\documentclass[aps,prd,twocolumn,nofootinbib,superscriptaddress,preprintnumbers]{revtex4-1}
\usepackage{amsfonts,amsmath,amssymb,amsthm,braket,nicefrac}
\usepackage{booktabs,array}
\usepackage[utf8]{inputenc}
\usepackage{float,graphicx,hhline}
\usepackage{algorithm,algpseudocode}
\usepackage{appendix}
\usepackage[dvipsnames]{xcolor}
\usepackage{mathtools}
\usepackage{multirow}
\usepackage{tikz-cd}
\usepackage{slashed}
\usepackage{soul}
\usepackage{pifont}
\usepackage[caption=false]{subfig}
\usepackage[export]{adjustbox}
\usepackage[normalem]{ulem}

\AtBeginDocument{
\heavyrulewidth=.08em
\lightrulewidth=.05em
\cmidrulewidth=.03em
\belowrulesep=.65ex
\belowbottomsep=0pt
\aboverulesep=.4ex
\abovetopsep=0pt
\cmidrulesep=\doublerulesep
\cmidrulekern=.5em
\defaultaddspace=.5em
}

\usepackage{hyperref}% add hypertext capabilities
\hypersetup{
    pdftitle={Flow-based sampling for fermionic lattice field theories},
    colorlinks=true,     % false: boxed links; true: colored links
    linkcolor=blue,      % color of internal links
    citecolor=blue,      % color of links to bibliography
    filecolor=blue,      % color of file links
    urlcolor=blue        % color of external links
}
\usepackage{cleveref}

\DeclareMathAlphabet{\mathbbold}{U}{bbold}{m}{n}
\newcommand{\pf}{\mathrm{pf}}
\newcommand{\KL}{\mathrm{KL}}

\DeclareMathOperator{\Tr}{Tr}
\DeclareMathOperator{\sign}{sign}

\begin{document}

\title{Flow-based sampling for fermionic lattice field theories}

\newcommand{\getMITAffiliation}{\affiliation{Center for Theoretical Physics, Massachusetts Institute of Technology, Cambridge, MA 02139, U.S.A.}}
\newcommand{\getNYUAffiliation}{\affiliation{Center for Cosmology and Particle Physics, New York University, New York, NY 10003, US}}
\newcommand{\getDMAffiliation}{\affiliation{DeepMind, London, UK}}
\newcommand{\getIAIFIAffiliation}{\affiliation{The NSF AI Institute for Artificial Intelligence and Fundamental Interactions}}

\author{Michael~S.~Albergo}
\email{albergo@nyu.edu}
\getNYUAffiliation
\author{Gurtej~Kanwar}
\email{gurtej@mit.edu}
\getMITAffiliation
\getIAIFIAffiliation
\author{S\'{e}bastien~Racani\`{e}re}
\email{sracaniere@google.com}
\getDMAffiliation
\author{Danilo~J.~Rezende}
\email{danilor@google.com}
\getDMAffiliation
\author{Julian~M.~Urban}
\email{urban@thphys.uni-heidelberg.de}
\affiliation{Institut f\"ur Theoretische Physik, Universit\"at Heidelberg, Philosophenweg 16, 69120 Heidelberg, Germany}
\author{\\Denis~Boyda}
\affiliation{Argonne Leadership Computing Facility, Argonne National Laboratory, Lemont IL-60439, USA}
\getMITAffiliation
\getIAIFIAffiliation
\author{Kyle~Cranmer}
\getNYUAffiliation
\author{Daniel~C.~Hackett}
\getMITAffiliation
\getIAIFIAffiliation
\author{Phiala~E.~Shanahan}
\getMITAffiliation
\getIAIFIAffiliation

\preprint{MIT-CTP/5307}

\begin{abstract}
    Algorithms based on normalizing flows are emerging as promising machine learning approaches to sampling complicated probability distributions in a way that can be made asymptotically exact. In the context of lattice field theory, proof-of-principle studies have demonstrated the effectiveness of this approach for scalar theories, gauge theories, and statistical systems. This work develops approaches that enable flow-based sampling of theories with dynamical fermions, which is necessary for the technique to be applied to lattice field theory studies of the Standard Model of particle physics and many condensed matter systems. As a practical demonstration, these methods are applied to the sampling of field configurations for a two-dimensional theory of massless staggered fermions coupled to a scalar field via a Yukawa interaction.
\end{abstract}
\maketitle

\date{\today}

\section{Introduction}

Lattice field theory is among the most successful approaches for regularizing and computing path integral expectation values in quantum field theory. In particular, the path integral can be numerically evaluated by formulating a stochastic process weighted by the Euclidean lattice action and applying Markov Chain Monte Carlo (MCMC) sampling~\cite{Morningstar:2007zm}. The Euclidean lattice results can then be systematically related to the corresponding continuum Minkowski theory. This procedure enables the investigation of equilibrium properties in the theory of interest, and is a powerful and well-established method to study strongly coupled quantum field theories non-perturbatively. Key areas of application include fundamental interactions, most prominently quantum chromodynamics (QCD), as well as problems in condensed matter theory; see Refs.~\cite{Brower:2019oor,Lehner:2019wvv,Kronfeld:2019nfb,Cirigliano:2019jig,Detmold:2019ghl,Bazavov:2019lgz,Mathur:2016cko,Joo:2019byq} for recent reviews.

The sequential nature of the Markov chain is a potential drawback to the MCMC sampling approach for computing path integrals in lattice field theory. In particular, known Markov chain update schemes for many theories of interest are local or diffusive, which can result in severe autocorrelations between successive elements of the chain. Naturally, the larger the autocorrelation between samples, the more samples must be drawn to achieve a result at fixed statistical precision. Close to criticality, e.g.~when approaching the continuum limit of lattice field theories, autocorrelations also diverge rapidly for such local or diffusive Markov chains. This issue, referred to as critical slowing down, can render computations prohibitively expensive~\cite{Wolff:1989wq,Schaefer:2009xx,Schaefer:2010hu}.

These challenges have motivated extensive work to replace local/diffusive MCMC algorithms, such as Hybrid Monte Carlo (HMC)~\cite{DUANE1987216}, with other sampling procedures. Specialized Markov chain steps have been developed in a number of specific contexts, including cluster updates~\cite{Hoshen:1976zz,Wolff:1987jk,Swendsen:1987ce,Edwards:1988ba,Wolff:1988uh,BrowerTamayo89,Hasenbusch:1989kx,Sinclair:1992vm,Bietenholz:1995zk}, worm algorithms~\cite{Evertz:2000rk,Prokofev:2001ddj,Kawashima:2004}, sampling in terms of dual variables~\cite{Savit:1979ny,Azcoiti:2009md,Gattringer:2015baa}, and event-chain algorithms~\cite{Bernard:2009,Michel:2015,Nishikawa:2015,Hasenbusch:2018ztj,Lei:2018vis}. Though these methods have been shown to mitigate critical slowing down in some settings, they cannot be applied to many theories of interest, including lattice QCD. In this light, development of sampling algorithms for lattice field theory based on machine learning techniques is underway, and previous works have applied a variety of tools such as adversarial learning and self-learning Monte Carlo methods~\cite{PhysRevD.100.011501,Wu_2019,Pawlowski:2018qxs,Nicoli:2020evf,Carrasquilla:2020mas,wang2020continuousmixture,vielhaben2020generative, PhysRevB.95.041101,PhysRevB.95.241104,PhysRevB.97.205140,PhysRevB.96.161102,PhysRevB.97.205140,PhysRevB.98.041102,PhysRevB.102.041124,Nagai:2020jar}. Progress has recently been made in substituting the proposal mechanism in MCMC with a variational ansatz based on a class of samplers known as normalizing flows~\cite{rezende2016variational,dinh2017density,JMLR:v22:19-1028}, which can be optimized to approximately sample from the target Boltzmann distribution~\cite{Albergo:2019eim,Nicoli:2020njz,DelDebbio:2021qwf}. Within this approach, proposed samples are by construction uncorrelated, and asymptotic exactness can be guaranteed by implementing a Markov chain with a Metropolis accept/reject step or through reweighting.

Though flow-based models have been extended to exactly incorporate the gauge symmetries inherent in many quantum field theories~\cite{Kanwar:2020xzo,Boyda:2020hsi}, existing applications are so far restricted to purely bosonic theories. For theories involving fermions, the anti-commuting nature of the associated operators must be treated by formulating the theory in terms of integrals over Grassmann-valued field variables. These integrals may be evaluated analytically, resulting in a purely bosonic theory described by an effective action which incorporates the dynamics of the fermion fields via `fermion determinant' terms. While flow-based methods may in principle be applied to learn this effective action over bosonic fields, the cost of computing such determinants scales unfavorably with the number of fermionic degrees of freedom, and their exact evaluation is typically intractable at the scale of state-of-the-art calculations.

\looseness=1
In this paper, a framework is presented for the application of flow-based sampling algorithms to lattice quantum field theories with dynamical fermions, such as lattice QCD and theories describing many condensed matter systems. We construct approaches based primarily on the pseudofermion method~\cite{Weingarten:1980hx} to avoid an explicit computation of the fermion determinant while guaranteeing asymptotic exactness of the sampling schemes. We investigate the application of a number of flow-based samplers in the context of a simple, two-dimensional Yukawa model with a real scalar field coupled to staggered fermions. Our results demonstrate that lattice field theories with dynamical fermions are amenable to flow-based sampling and provide a starting point for extensions to higher-dimensional settings as well as theories involving gauge fields. The architectures developed here are also applicable to flow-based acceleration of traditional MCMC approaches~\cite{song2018anicemc,levy2018generalizing,medvidovic2021generative,Foreman:2021ixr,gabrie2021adaptive}. The primary contributions of this work are:
\begin{itemize}
    \item In \Cref{sec:sampling-dists}, identifying four distinct sampling schemes based on generative models that capture different decompositions/marginalizations of the target distribution over boson and pseudofermion field configurations;
    \item In \Cref{sec:fermion-flows}, constructing and optimizing efficient, expressive flow model architectures that respect the symmetries of the pseudofermion action, in particular translational symmetry with antiperiodic temporal boundary conditions;
    \item In \Cref{sec:results}, implementing and numerically benchmarking these sampling approaches in the context of a two-dimensional field theory with one pair of mass-degenerate fermions.
\end{itemize}

\looseness=1
The remainder of this paper is organized as follows. In \Cref{sec:theory}, we review the description of fermions in lattice field theory, the use of pseudofermions, and the boundary conditions and translational symmetry of pseudofermion actions. In \Cref{sec:sampling-dists}, we outline four exact generative sampling schemes for fermionic theories and subsequently develop suitable flow architectures as the generative models for use in these sampling schemes in \Cref{sec:fermion-flows}. Details and numerical results of the application of our framework to a Yukawa theory in two dimensions are presented in \Cref{sec:results}. In \Cref{sec:update}, we discuss the applicability of our developments to update-based approaches. We summarize our findings and provide an outlook in \Cref{sec:summary}.

\section{Fermions on the lattice}
\label{sec:theory}

The simulation of dynamical fermion degrees of freedom in lattice field theory is a highly non-trivial task for many theories of physical interest, both conceptually and computationally. In this section, we briefly review the main concepts behind formulations of lattice fermions and their numerical implementation. For a comprehensive treatment we refer the reader to one of the standard textbooks; see e.g.~Refs.~\cite{Gattringer:2010zz,Montvay:1994cy,Rothe:1992nt}.

\subsection{Path integrals with fermions}

We consider field theories of interacting fermionic and bosonic degrees of freedom discretized on a $d$-dimensional Euclidean hypercubic lattice with periodic boundary conditions. The action of such a theory can be expressed as
\begin{equation}\label{eq:fermionic-theory-action}
    S(\psi,\bar{\psi},\phi) = S_B(\phi) + S_F(\psi,\bar{\psi},\phi)\ ,
\end{equation}
where the subscripts $B$ and $F$ denote the bosonic and fermionic contributions to the action, the discretized boson field variables are collectively denoted by $\phi$, and the discretized fermion field variables are denoted by $\psi,\bar{\psi}$. For the present work, we assume that the fermionic action is bilinear in $N_f$ flavors of Dirac fermions $\psi_f,\bar{\psi}_f$ and is given by
\begin{equation}\label{eq:fermion-action}
    S_F(\psi,\bar{\psi},\phi) = \sum_{f=1}^{N_f} \bar{\psi}_f \, D_f(\phi)\, \psi_f \ ,
\end{equation}
where the Dirac operator $D_f(\phi)$ includes the kinetic terms, mass terms, and any coupling to bosonic fields for each fermion flavor $f$. The precise form of $D_f(\phi)$ is determined by the theory of interest and the choice of discretization.

Expectation values of observables $\mathcal{O}$ are computed via path integrals of the form
\begin{equation}\label{eq:fermions_integral}
    \braket{\mathcal{O}}  = \frac{1}{Z} \int \mathcal{D}[\phi] \mathcal{D}[\psi, \bar{\psi}] e^{-S_F(\psi,\bar{\psi}, \phi)} e^{-S_B(\phi)} \mathcal{O}(\psi, \bar{\psi}, \phi)\ ,
\end{equation}
where
\begin{equation}\label{eq:normalizer-Z}
    Z = \int \mathcal{D}[\phi] \mathcal{D}[\psi, \bar{\psi}]  \, e^{-S_F(\psi,\bar{\psi},\phi)}e^{-S_B(\phi)}\ ,
\end{equation}
and the fermion fields $\psi$ and $\bar{\psi}$ are defined in terms of anti-commuting Grassmann numbers. For bilinear actions of the form given in \Cref{eq:fermion-action}, integration over the Grassmann-valued fermion fields can be performed explicitly, giving
\begin{equation}
    \int \mathcal{D}[\psi, \bar{\psi}] e^{-S_F(\psi,\bar{\psi}, \phi)} = \prod_{f=1}^{N_f} \det{D_f(\phi)}\ .
\end{equation}
By applying Wick's theorem, the dependence of the observable on the fermions can be integrated out. Path integral expectation values can then be written in terms of purely bosonic degrees of freedom as
\begin{equation}\label{eq:bosonic-obs-eval}
    \braket{\mathcal{O}} = \frac{1}{Z} \int \mathcal{D}[\phi] \left[\prod_{f=1}^{N_f} \det{D_f(\phi)}\right] e^{-S_B(\phi)} \mathcal{O}(\phi)\ .
\end{equation}
This expectation value may be estimated via MCMC sampling by computing an average over a statistical ensemble of configurations $\phi$,
\begin{equation}
    \langle\mathcal{O}\rangle \approx \frac{1}{N} \sum_{\phi \sim p} \mathcal{O}(\phi)\ ,
\end{equation}
where $\sum_{\phi \sim p}$ denotes a sum over $N$ configurations sampled from the probability distribution
\begin{equation}\label{eq:marginal-dist}
    p(\phi) = \frac{1}{Z} e^{-S_B(\phi)} \prod_{f=1}^{N_f} \det D_f(\phi)\ .
\end{equation}

Direct sampling schemes for high-dimensional lattice distributions are typically not known, even for theories without fermions. Nevertheless, the distribution $p(\phi)$ can be sampled via MCMC methods with guaranteed asymptotic exactness under certain ergodicity and balance constraints~\cite{meyn2012markovErgodicity}. Among these, the HMC algorithm~\cite{DUANE1987216} has been established as the de facto standard method for producing configurations in lattice field theory and is routinely employed in state-of-the-art QCD studies and beyond. This algorithm is based on the numerical treatment of Hamiltonian equations of motion in a fictitious time dimension, where quantum fluctuations are encoded by the random sampling of the associated canonical momenta. Given a field configuration and a set of momenta, this evolution is computed with a symplectic integrator such as the leapfrog algorithm. A Metropolis-Hastings accept/reject step~\cite{Metropolis:1953am,Hastings:1970aa} results in an algorithm satisfying detailed balance, despite the accumulation of numerical errors along the discretized integration trajectory. At scale, the fermion determinants are treated stochastically via the pseudofermion method introduced in the following section.

\subsection{Pseudofermions}
\label{sec:pseudofermions}

The fermion determinants in \Cref{eq:marginal-dist} cannot be calculated directly at scale because the Dirac matrices $D_f$ are high-dimensional. For $d$-dimensional field configurations with $L$ sites per spatial dimension and $L_t$ sites in the temporal dimension, the total number of fermionic degrees of freedom scales as the total number of lattice sites, $V = L_t L^{d-1}$. Each Dirac matrix $D_f$ then has dimensions $\mathcal{O}(V \times V)$, and an exact computation of the determinants becomes intractable at the scales of many theories of interest.

Instead, Gaussian integrals over auxiliary bosonic fields can be used to replace the direct evaluation of determinant factors, based on the identity
\begin{equation}\label{eq:pf-determinant}
    \det \mathcal{M} = \frac{1}{Z_{\mathcal{N}}} \int \mathcal{D}[\varphi_R,\varphi_I] \, e^{-\varphi^\dagger \mathcal{M}^{-1} \varphi}\ ,
\end{equation}
where the normalization constant $Z_\mathcal{N}$ is defined as
\begin{equation}\label{eq:gauss-normalization}
    Z_{\mathcal{N}} = \int \mathcal{D}[\varphi_R,\varphi_I] \, e^{-\varphi^\dagger \varphi}\ .
\end{equation}
Here, $\varphi_R,\varphi_I$ denote the real and imaginary components of the auxiliary complex field $\varphi$, and the matrix $\mathcal{M}$ must be positive-definite. Since the Dirac matrices $D_f$ are typically not positive-definite, one cannot directly apply this identity to each factor of $\det{D_f}$. However, for fermion flavors $f_1$ and $f_2$ appearing as degenerate pairs, based on $\gamma_5$-hermiticity one can instead use the equality
\begin{equation}
    \det{D_{f_1}} \det{D_{f_2}} = \det{D_{f_1} D_{f_1}^\dagger}\ ,
\end{equation}
and then apply \Cref{eq:pf-determinant} to the positive-definite matrix ${\mathcal{M} = D_{f_1} D_{f_1}^\dagger}$. For fermion flavors $f$ not included in any degenerate pair, one can apply one-flavor algorithms~\cite{Kennedy:1998cu,Ogawa:2009ex,Chen:2014hyy} to replace $D_f$ with a positive-definite matrix $\mathcal{M}$ capturing identical dynamics.

Using the pseudofermion approach, a path integral as in \Cref{eq:bosonic-obs-eval} can thus be rewritten in terms of an action involving the auxiliary pseudofermion fields $\varphi$,
\begin{equation}\label{eq:pseudo-action}
\begin{aligned}
    S(\phi,\varphi) &= S_{B}(\phi) + S_{PF}(\phi,\varphi) \ \ \text{with}\ \\
    S_{PF}(\phi,\varphi) &= \varphi^\dag \mathcal{M}^{-1}(\phi) \varphi \equiv \sum_{k=1}^{N_{\pf}} \varphi_k^\dag \mathcal{M}_k^{-1}(\phi) \varphi_k \ ,
\end{aligned}
\end{equation}
after replacing the fermion determinants in the given lattice theory by the determinants of $N_{\pf}$ positive-definite matrices $\mathcal{M}_k$ as
\begin{equation}
    \prod_{f=1}^{N_{f}} \det D_f(\phi) = \prod_{k=1}^{N_{\pf}} \det \mathcal{M}_k(\phi)\ .
\end{equation}
Each term $\varphi^{\dagger}_k \mathcal{M}_k^{-1} \varphi_k$ in the pseudofermion action can be efficiently computed using iterative solvers such as the conjugate gradient method. Having formulated the theory using pseudofermions in \Cref{eq:pseudo-action}, evaluation of the path integral via MCMC can then be performed in this augmented space by sampling from the joint distribution
\begin{equation}\label{eq:joint-dist}
    p(\phi,\varphi) = \frac{1}{Z} e^{-S_{B}(\phi) - S_{PF}(\phi,\varphi)}\ .
\end{equation}
The distribution $p(\phi)$ given in \Cref{eq:marginal-dist} is obtained from this joint distribution by marginalizing over the auxiliary pseudofermion fields $\varphi$.

\begin{table*}
    \renewcommand{\arraystretch}{2.0}
    \begin{ruledtabular}
    \begin{tabular}{>{\centering\arraybackslash}m{2.5cm} >{\hfill\arraybackslash}m{3.0cm} m{7.5cm} >{\centering\arraybackslash}m{4.5cm}}
        \centering
        Name & \multicolumn{2}{c}{Probability density} & Use case \\
        \midrule
        Joint${}^\text{A}$ & $p(\phi, \varphi)$ & $=\quad \frac{1}{Z} \exp(-S_{B}(\phi)-\varphi^\dagger \left[\mathcal{M}(\phi)\right]^{-1} \varphi)$ &  \Cref{sec:joint-sampling} \\ 
        $\phi$-marginal & $p(\phi)$ & $=\quad \frac{Z_{\mathcal{N}}}{Z} \exp(-S_{B}(\phi)) \det \mathcal{M}(\phi)$ & \Cref{sec:phi-marginal-sampling,sec:autoregressive-sampling}\\ 
        $\varphi$-conditional${}^{\text{A},\text{B}}$ & $p(\varphi | \phi)$ & $=\quad \dfrac{1}{Z_{\mathcal{N}} \det{\mathcal{M}(\phi)}} \exp(-\varphi^\dagger \left[\mathcal{M}(\phi)\right]^{-1} \varphi)$  & \Cref{sec:phi-marginal-sampling,,sec:gibbs-sampling,,sec:autoregressive-sampling}  \\ 
        $\varphi$-marginal${}^\text{C}$ & $p(\varphi)$ & $=\quad \frac{1}{Z} \int d\phi\, \exp(-S_B(\phi) -\varphi^\dagger \left[\mathcal{M}(\phi)\right]^{-1} \varphi)$ & -- \\ 
        $\phi$-conditional${}^\text{A}$ & $p(\phi|\varphi)$ & $=\quad \dfrac{\exp(-S_{B}(\phi) -\varphi^\dagger \left[\mathcal{M}(\phi) \right]^{-1} \varphi)}{\int d\phi \, \exp(-S_{B}(\phi) - \varphi^\dagger \left[\mathcal{M}(\phi) \right]^{-1} \varphi)}$ & \Cref{sec:gibbs-sampling}\\[3ex]
    \end{tabular}
    \end{ruledtabular}
    \caption{List of possible distributions derived from the joint target density in \Cref{eq:joint-dist}. The normalizing constant $Z$ is given by \Cref{eq:normalizer-Z} and $Z_{\mathcal{N}}$ is defined in \Cref{eq:gauss-normalization}. Notes: (A) Only the joint, $\varphi$-conditional, and $\phi$-conditional densities can be efficiently computed (up to normalization). (B) The $\varphi$-conditional can be sampled exactly by the method specified in \Cref{eq:pf-sampling}. (C) A closed form for the $\varphi$-marginal density is not generally known (even unnormalized).}
    \label{tab:dists}
\end{table*}

While the pseudofermion method renders the treatment of fermion determinants tractable in principle, the joint action may strongly fluctuate in certain limits. This feature can slow down MCMC sampling of the joint distribution and can lead to an unfavorable volume scaling of the associated computational effort, especially when many components of the bosonic field are updated simultaneously. Accordingly, numerous modifications of the pseudofermion formulation have been developed to improve the structure of the action; see e.g.~Refs.~\cite{deForcrand:1996ck,deForcrand:1998sv,Peardon:2000si,Hasenbusch:2001xh,Clark:2006wq,Clark:2006fx}. These developments are complementary to the application of generative models for sampling the pseudofermion distribution and could be combined with any of the approaches presented here. For example, in this work we improve the efficiency of modeling and sampling the distributions under study by applying even-odd preconditioning~\cite{DeGrand:1990dk}, as discussed in \Cref{sec:yukawa-theory} and \Cref{app:even-odd}.

\subsection{Boundary conditions and translational symmetry}
\label{sec:fermions-sym}

In lattice studies of purely bosonic theories, it is common to choose periodic boundary conditions in all directions of the lattice, allowing the incorporation of an exact discrete translational symmetry in lattice actions for such theories. For fermion fields, one needs to impose antiperiodic boundary conditions in the time direction in order to obtain a consistent definition of the trace for the Euclidean partition function. Actions for theories involving fermionic fields are then invariant under simultaneous spatio-temporal translations of $\phi$, $\psi$, and $\bar\psi$, with appropriate boundary conditions applied for each field.

To be consistent with the boundary conditions for $\psi$ and $\bar{\psi}$, each Dirac matrix $D_f(\phi)$ must include appropriate signs for any terms coupling fields across the temporal boundary. As a result, these boundary conditions affect the pseudofermion formulation of the theory as well, and the pseudofermion action $S_{PF}(\phi,\varphi)$ is invariant under simultaneous translations of $\phi$ and $\varphi$ with antiperiodic temporal boundary conditions applied to $\varphi$.

\begin{figure*}
    \centering
    \subfloat[$\phi$-Marginal (\Cref{sec:phi-marginal-sampling})\label{fig:marginal}]{%
        \includegraphics[valign=c]{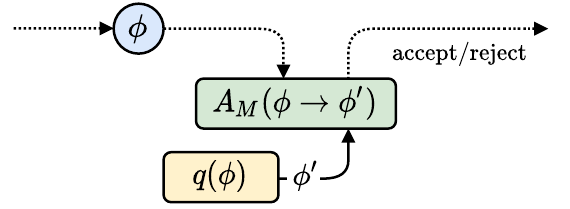}}%
    \hspace{5ex}
    \subfloat[Gibbs (\Cref{sec:gibbs-sampling})\label{fig:Gibbs}]{%
        \includegraphics[valign=c]{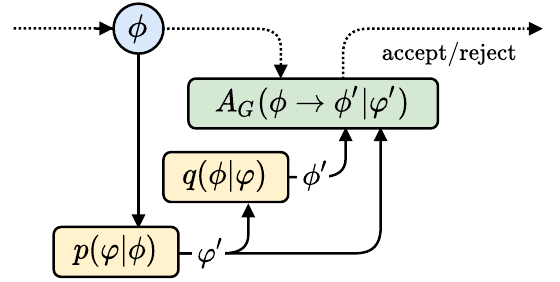}}
    \vspace{5ex}
    \subfloat[Autoregressive (\Cref{sec:autoregressive-sampling})\label{fig:autoregressive}]{%
        \includegraphics[valign=c]{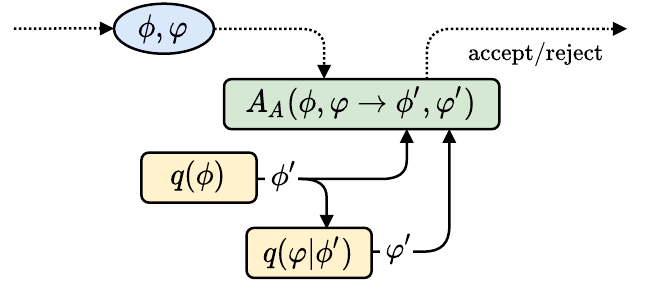}}%
    \hspace{5ex}
    \subfloat[Joint (\Cref{sec:joint-sampling})\label{fig:joint}]{%
        \includegraphics[valign=c]{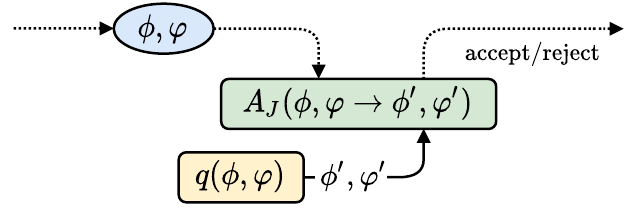}}%
    \caption{Diagrams illustrating the four types of sampling schemes described in \Cref{sec:sampling-dists}. Blue circles/ellipses depict the current state of the Markov chain. Yellow boxes depict exactly sampleable densities either produced from generative models or by \Cref{eq:pf-sampling}. Green boxes correspond to Metropolis accept/reject steps using the acceptance probabilities defined in the text. Dotted lines indicate the Markov chain, whereas solid lines correspond to the internal operations of each Markov chain step.}
    \label{fig:exact-mcmc}
\end{figure*}

In general, the discretization chosen for the Dirac operator determines which particular lattice translations are included in the translational symmetry group. In the staggered formulation~\cite{Kogut:1974ag}, for example, the spinor components of each flavor of fermion are distributed over the components of hypercubes with $2^{d}$ sites each, and the translational symmetry group includes all translations by an even number of sites. Translations by an odd number of sites in particular directions correspond to more complicated internal symmetry transformations that mix spinor degrees of freedom, and must involve sign flips on specific field components to leave the staggered action invariant~\cite{Kronfeld:2007ek}. These translational symmetries of the staggered formulation play a role in the application to the staggered-fermion Yukawa model presented in \Cref{sec:results}.

\section{Exact generative sampling schemes for fermionic theories}
\label{sec:sampling-dists}

Generating importance-weighted field configurations for a lattice field theory involving fermions can proceed via the marginal distribution $p(\phi)$ defined in \Cref{eq:marginal-dist}, the joint distribution $p(\phi,\varphi)$ defined in \Cref{eq:joint-dist}, or through other choices of marginalized distributions defined in \Cref{tab:dists}. In this work, we develop exact sampling schemes based on generative models that directly approximate these distributions. In defining these sampling schemes, we assume that the model probability density may be computed for each generative model (this property holds for the flow-based models defined below). This section details four asymptotically exact schemes for constructing Markov chains to draw samples of $\phi$, as illustrated in \Cref{fig:exact-mcmc}.

\subsection{Modeling and sampling of \texorpdfstring{$p(\phi)$}{p(phi)}}
\label{sec:phi-marginal-sampling}

Since we must ultimately sample only the field $\phi$, one could directly model the $\phi$-marginal distribution (Row 2 of \Cref{tab:dists}) by constructing a generative sampler providing a distribution $q(\phi)$ approximating $p(\phi)$. Samples drawn from the model distribution $q(\phi)$ can be used in asymptotically exact sampling schemes by either constructing an independence Metropolis Markov chain or applying reweighting/resampling based on reweighting factors $p(\phi)/q(\phi)$. A direct application of either approach requires computing $p(\phi)$ involving the aforementioned determinant factors. The acceptance probability for such a Metropolis Markov chain would be
\begin{equation}\label{eq:marginal-acc-exact}
    A_M(\phi \rightarrow \phi') = \min\left(1, \frac{e^{-S_B(\phi')} \det \mathcal{M}(\phi')}{e^{-S_B(\phi)} \det \mathcal{M}(\phi)}\frac{q(\phi)}{ q(\phi')}\right)\ .
\end{equation}
This sampling scheme is illustrated in \Cref{fig:marginal}. 

Instead of evaluating the ratio $\det \mathcal{M}(\phi') / \det \mathcal{M}(\phi)$ directly, which becomes prohibitive at scale, it is possible to apply the pseudo-marginal method~\cite{Andrieu:2009} using stochastic approximations of both the numerator and denominator of \Cref{eq:marginal-acc-exact} in a way that retains asymptotic exactness. In this stochastic generalization of the Metropolis algorithm, one computes an estimate of $p(\phi)$ using an unbiased stochastic estimator when $\phi$ is initially proposed. This estimate of $p(\phi)$ is then used in all subsequent accept/reject tests for the next element in the Markov chain. Applied to a theory with fermions, this amounts to computing a stochastic estimate of the fermion determinant for each proposed configuration.\footnote{Note that the pseudo-marginal algorithm is not equivalent to the sampling scheme based on stochastic estimates of ratios~\cite{Gattringer:2010zz,Bhanot:1983uy}, for which asymptotic exactness has not been demonstrated.}

For example, we can use an unbiased estimator based on pseudofermions. An (unnormalized) estimate for $p(\phi)$ can be obtained by generating a pseudofermion $\varphi$ from the conditional $p(\varphi|\phi)$ and measuring the quantity $e^{-\varphi^\dag (\mathcal{M}^{-1}(\phi) - 1) \varphi} e^{-S_B(\phi)}$. The $\varphi$-conditional can be directly sampled according to
\begin{equation}\label{eq:pf-sampling}
    \varphi = \mathcal{A}(\phi) \chi, \quad \mathrm{where} \quad \chi \sim \frac{1}{Z_{\mathcal{N}}}e^{-\chi^\dag \chi}\ .
\end{equation}
The matrix $\mathcal{A}$ is defined by the identity ${\mathcal{M}(\phi) \equiv \mathcal{A}(\phi) \mathcal{A}^\dag(\phi)}$ and reduces to the Dirac matrix $D(\phi)$ in a two-flavor example. This estimate can be extremely noisy in practice and may give poor statistical performance. However, it can be improved upon by using multiple pseudofermion draws; see e.g.~Refs.~\cite{hutchinson1990,Hasenbusch:2001xh,Clark:2006fx,deForcrand:2018orx}.

In principle, any unbiased stochastic estimator of the fermion determinant can be applied (whether based on pseudofermions or entirely distinct). The limit of taking arbitrarily precise estimators recovers the exact acceptance probability in \Cref{eq:marginal-acc-exact}. Acceptance rates obtained by using the exact form can thus be interpreted as an upper bound on sampling performance.

\subsection{Gibbs sampling using \texorpdfstring{$p(\phi|\varphi),p(\varphi|\phi)$}{p(phi|varphi), p(varphi|phi)}}
\label{sec:gibbs-sampling}

An alternative to modeling $p(\phi)$ directly is to construct samplers for both conditional distributions $p(\phi|\varphi)$ and $p(\varphi|\phi)$ and build an asymptotically exact Gibbs sampler that alternatingly samples from these distributions to update $\phi$ and $\varphi$. For such a Gibbs sampler to satisfy detailed balance, the update to $\phi$ must satisfy detailed balance for $p(\phi|\varphi)$ and the update to $\varphi$ must satisfy detailed balance for $p(\varphi|\phi)$. The $\varphi$-conditional can be exactly and directly sampled as described in \Cref{eq:pf-sampling}, automatically fulfilling this requirement. On the other hand, the $\phi$-conditional (Row 5 of \Cref{tab:dists}) can be approximated by a generative model distribution $q(\phi|\varphi) \approx p(\phi|\varphi)$. This model can be incorporated into an exact Markov chain transition for the $\phi$-conditional distribution as follows. Start with a state $\phi$, sample $\varphi'$ from $p(\varphi|\phi)$ using \Cref{eq:pf-sampling}, conditionally propose $\phi'$ from $q(\phi|\varphi')$, and then apply a Metropolis-Hastings accept/reject step with the acceptance probability given by
\begin{equation}\label{eq:gibbs-acc}
\begin{aligned}
    A_G(\phi \rightarrow \phi'|\varphi') &= \min\left(1, \frac{p(\phi'|\varphi')}{p(\phi|\varphi')}\frac{q(\phi|\varphi')}{q(\phi'|\varphi')} \right) \\
    &= \min\left(1, \frac{p(\phi',\varphi')}{p(\phi,\varphi')}\frac{q(\phi|\varphi')}{q(\phi'|\varphi')} \right)\ .
\end{aligned}
\end{equation}
This step satisfies detailed balance for the $\phi$-conditional distribution $p(\phi|\varphi)$ as required and guarantees asymptotic exactness. Note that in contrast to the $\phi$-marginal sampler described in \Cref{sec:phi-marginal-sampling}, computing the acceptance probability at scale for the sampling scheme described here and in \Cref{sec:autoregressive-sampling,sec:joint-sampling} does not rely on unbiased stochastic determinant estimators.

In this approach, the field $\varphi'$ is independently re-sampled conditioned on $\phi$ at each step of the Markov chain, and therefore does not need to be stored. This Gibbs sampler can thus be interpreted as an exact Markov chain over $\phi$ alone, with the sampling of $\varphi'$ contained inside each Markov chain step as depicted in \Cref{fig:Gibbs}. The approach closely mirrors the typical sampling strategy employed in HMC, in which pseudofermions $\varphi'$ are sampled according to the exact conditional distribution $p(\varphi|\phi)$ and Hamiltonian evolution is used to construct an update step that satisfies detailed balance for the conditional distribution $p(\phi|\varphi')$. The generative model proposal and Metropolis-Hastings step for $p(\phi|\varphi')$ can thus be considered an optimizable replacement of the molecular dynamics trajectory utilized in HMC, with the difference that the mechanism of generating a proposal configuration $\phi'$ does not directly depend on $\phi$ (as is the case for a symplectic integrator), but only indirectly through $\varphi'$. However, this also means that in contrast to all other schemes described here, the Gibbs sampler is not an independence sampler. Drawing configurations from the model and constructing the Markov chain cannot be done asynchronously, since the generation of a proposal explicitly depends on the previous element of the chain.

\subsection{Autoregressive modeling and sampling of \texorpdfstring{$p(\phi,\varphi)$}{p(phi,varphi)}}
\label{sec:autoregressive-sampling}

The joint distribution $p(\phi,\varphi)$ can be autoregressively decomposed as the product $p(\phi,\varphi) = p(\phi) p(\varphi|\phi)$ in terms of the $\phi$-marginal and $\varphi$-conditional (Rows 2 and 3 of \Cref{tab:dists}). A generative model for the joint distribution could therefore be produced by approximating both components independently, i.e., $q(\phi,\varphi) = q(\phi) q(\varphi|\phi)$. This autoregressive decomposition allows the joint distribution to be reproduced in terms of two potentially simpler distributions. Note that although the exact sampling procedure described in \Cref{eq:pf-sampling} can be applied to draw samples from $p(\varphi|\phi)$, computing the normalizing constant of this $\varphi$-conditional distribution is not tractable. This is not an obstacle when one is only interested in conditionally sampling $\varphi$, as is the case for HMC or the approaches of \Cref{sec:phi-marginal-sampling,sec:gibbs-sampling}, but motivates modeling the distribution in the case where an approximation with a tractable density is required.

Exactness can be straightforwardly enforced in this approach by employing Markov chain steps in which joint samples $(\phi',\varphi')$ are proposed independently from $q(\phi,\varphi)$, and a Metropolis-Hastings accept/reject step is applied for the proposed transition $(\phi,\varphi) \rightarrow (\phi',\varphi')$ according to the acceptance probability
\begin{equation}\label{eq:autoregressive-acc}
    A_A(\phi,\varphi \rightarrow \phi',\varphi') = \min\left(1, \frac{p(\phi',\varphi')}{p(\phi,\varphi)}\frac{q(\phi)q(\varphi|\phi)}{q(\phi')q(\varphi'|\phi')} \right)\ .
\end{equation}
This sampling scheme is illustrated in \Cref{fig:autoregressive}. Furthermore, unique reweighting factors can be tractably computed for each configuration $\phi$ as $p(\phi,\varphi) / q(\phi)q(\varphi|\phi)$, thus reweighting approaches may also be used as alternatives to MCMC in order to guarantee exactness here.

\subsection{Fully joint modeling and sampling of \texorpdfstring{$p(\phi,\varphi)$}{p(phi,varphi)}}
\label{sec:joint-sampling}

Rather than modeling the factors $p(\phi)$ and $p(\varphi|\phi)$, one could instead apply generative models to jointly sample the fields $\phi$ and $\varphi$ according to a distribution $q(\phi,\varphi)$ that directly approximates the joint distribution (Row 1 of \Cref{tab:dists}). This results in joint samples and density estimates analogous to the autoregressive case above, but is a qualitatively distinct approach to modeling this distribution. Exactness can be enforced using a similar Metropolis-Hastings Markov chain transition with acceptance probability
\begin{equation}\label{eq:joint-acc}
    A_J(\phi,\varphi \rightarrow \phi',\varphi') = \min\left(1, \frac{p(\phi',\varphi')}{p(\phi,\varphi)}\frac{q(\phi,\varphi)}{q(\phi',\varphi')} \right)\ ,
\end{equation}
or by applying reweighting or direct resampling techniques. This approach is illustrated in \Cref{fig:joint}.

\section{Fermionic flows via pseudofermions}
\label{sec:fermion-flows}

The sampling approaches discussed above for theories involving fermions can in principle use any generative models that enable both efficient sampling and density estimation for the relevant model distributions. Normalizing flows are one such class of probabilistic models for which these operations are made possible using a change-of-variables formula~\cite{tabak2010,tabak2013,rezende2016variational}. Flow-based methods have been successfully implemented to model the unnormalized Boltzmann distributions of $\phi^4$-theory as well as $U(1)$ and $SU(N)$ gauge theories~\cite{Albergo:2019eim,Nicoli:2020njz,Kanwar:2020xzo,Boyda:2020hsi}, opening up a wealth of potential applications in high energy and nuclear physics as well as condensed matter theory. For an in-depth introduction to normalizing flows for lattice field theory with further implementation details and explanations, we refer the interested reader to Ref.~\cite{Albergo:2021vyo}. For a general introduction to machine learning for physicists, we recommend Ref.~\cite{Mehta_2019}.

Below, we outline the key concepts behind flows relevant for our approach to modeling the distributions listed in \Cref{tab:dists}:
\begin{enumerate}
    \item A flow-based model consists of an invertible `flow' $f$ and a prior distribution with probability density $r(\cdot)$, which can together be used to produce samples by first drawing $z$ according to $r(z)$, then returning $x = f(z)$;\footnote{We use the generic notation $z$ and $x$ here to stand for potentially high-dimensional variables acted on by the flow. In our applications these may include boson and/or pseudofermion field variables. Flows are described in field-theoretic notation wherever we work with flows particular to lattice field theory sampling.}
    \item The Jacobian $J = \frac{\partial f}{\partial z}$, combined with the prior density, allows the output probability density $q$ to be evaluated as
    \begin{equation}\label{eq:logq-generic}
        q(x) = r(z) \left|\det \frac{\partial f}{\partial z} \right|^{-1}\ ;
    \end{equation}
    \item The flow $f$ is parameterized by free model parameters which may be optimized by minimizing a suitable `loss function' that quantifies the difference between the model distribution $q(x)$ and the target distribution. A common choice also employed in the present work is the Kullback-Leibler (KL) divergence \cite{kullback1951information}.
\end{enumerate}

The remainder of this section describes how models for each of the distributions required for sampling may be constructed. First, a common training procedure for all such models is described in \Cref{sec:flow-optim} based on the idea that each distribution aims to approximate some marginalization of the same joint distribution $p(\phi,\varphi)$. This common training procedure motivates some of the architectural decisions for the construction of the models described in \Cref{sec:building-blocks,sec:flow-models}. In the following, we label the model densities according to their corresponding target densities in \Cref{tab:dists} as $q(\phi,\varphi)$, $q(\phi)$, $q(\varphi|\phi)$, and $q(\phi|\varphi)$. In each sampling approach, using model distributions that better approximate the associated target will generally result in higher acceptance rates with potentially lower autocorrelations.

\subsection{Optimizing flow-based models}
\label{sec:flow-optim}

We first detail a procedure to optimize the model density $q(\phi, \varphi)$ to directly approximate $p(\phi, \varphi)$. The KL divergence between these distributions is defined as
\begin{equation}\label{eq:kl_joint}
\begin{aligned}
    &D_{\KL}(q(\phi, \varphi) || p(\phi, \varphi)) \\
    &\hspace{10pt} = \mathbb{E}_{\phi, \varphi \sim q}\left[\log(q(\phi, \varphi) / p(\phi, \varphi))\right] \\
    &\hspace{10pt} = \mathbb{E}_{\phi, \varphi \sim q}\left[ \log q(\phi,\varphi) + S_B(\phi) + S_{PF}(\phi,\varphi) + \log{Z} \right]\ .
\end{aligned}
\end{equation}
It is minimized if and only if the target and model density match, for which ${D_{\KL} = 0}$. In practice, a loss function based on this divergence is computed stochastically as
\begin{equation}\label{eq:kl_joint_estimate}
    L = \frac{1}{N}\sum_{k=1}^N\log q(\phi_k, \varphi_k) + S_B(\phi_k) + S_{PF}(\phi_k, \varphi_k)\ ,
\end{equation}
in terms of a mini-batch of $N$ samples $\{(\phi_k, \varphi_k)\}_{k=1}^{N}$ drawn from $q(\phi,\varphi)$. The unknown normalizing constant $\log Z$ has been removed in the definition of \Cref{eq:kl_joint_estimate}, since it is just an overall constant shift and does not affect the relevant structure of the loss function.

If the model probability density $q(\phi,\varphi)$ can be directly computed, we can evaluate the gradient of \Cref{eq:kl_joint_estimate} with respect to the model parameters defining this probability density. Gradient-based optimization methods can then be applied to minimize $L$. This training procedure is immediately applicable to the models required for the joint sampling approaches derived in \Cref{sec:autoregressive-sampling,sec:joint-sampling}. In the former, the distribution $q(\phi,\varphi)$ is defined by $q(\phi)q(\varphi|\phi)$, and this pair of model distributions is simultaneously optimized by minimizing the loss function in \Cref{eq:kl_joint_estimate}. In the latter, a model for $q(\phi,\varphi)$ is directly constructed and optimized.

The remaining distributions required in \Cref{sec:phi-marginal-sampling,sec:gibbs-sampling}, namely $q(\phi)$ and $q(\phi|\varphi)$, do not naturally define a joint model probability density. To optimize these distributions using the loss function above, we extend the model architectures by pairing $q(\phi | \varphi)$ with a sampler $q(\varphi)$ and pairing $q(\phi)$ with a sampler $q(\varphi | \phi)$. The resulting joint models can be optimized as above, and the auxiliary  components can be discarded after training. In this work, these auxiliary models are constructed as follows.

We first consider extending the $\phi$-conditional model $q(\phi|\varphi)$ to a joint model,  which requires a marginal distribution $q(\varphi)$. None of the sampling approaches presented in \Cref{sec:sampling-dists} directly require this marginal distribution; however, as we discuss further in \Cref{sec:flow-models}, we choose to model $q(\phi|\varphi)$ by a restricted form of a joint sampler which simultaneously models a marginal distribution $q(\varphi)$. In this extended model, both $q(\phi | \varphi)$ and $q(\varphi)$ are described by parameters that are optimized.

A $\phi$-marginal model $q(\phi)$ can be extended to a joint sampler by pairing it with a conditional distribution $q(\varphi|\phi)$. In principle such an auxiliary model could be constructed solely for the purposes of training. However, in this case we are free to instead use the exact conditional distribution $p(\varphi|\phi)$, which can be exactly and efficiently sampled. The result is a joint distribution defined by first sampling $\phi$ from the $\phi$-marginal model and then sampling the $\varphi$-conditional using \Cref{eq:pf-sampling}, resulting in the joint density ${q(\phi) p(\varphi|\phi)}$. Evaluating the joint KL divergence between this model distribution and the target joint distribution $p(\phi,\varphi)$ requires the evaluation of the normalized density $p(\varphi|\phi)$, which unfortunately includes a normalizing factor of $\det \mathcal{M}(\phi)$. However, we only require an unbiased stochastic estimator of the gradients of \Cref{eq:kl_joint_estimate} for optimization. \Cref{app:varphi-estimated-grad} details a particular stochastic estimator for these gradients which can be used to avoid the costly determinant evaluation; this estimator is used for all $\phi$-marginal models trained in this work.

\subsection{Building blocks}
\label{sec:building-blocks}

\begin{figure*}
    \centering
    \includegraphics{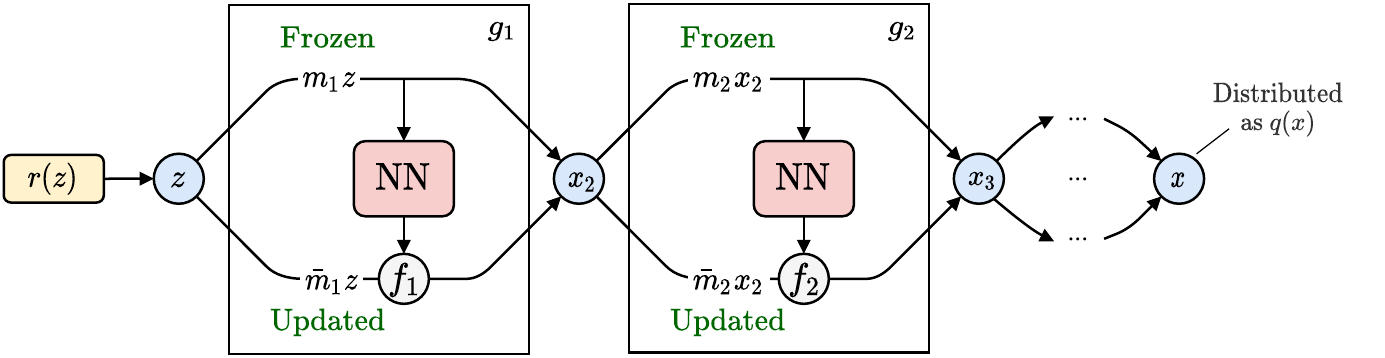}
    \caption{Schematic overview of a coupling-layer-based flow model architecture. Masks $m_i$ and their complements $\bar{m}_i = 1 - m_i$ split the degrees of freedom into a subset to be updated and a subset that is frozen and used as input to the context function (red box), which provide the parameters for the invertible transformation $f_i$ applied within each coupling layer $g_i$. Here, neural networks (NN) are used to implement the context functions. The generic variables $z$, $x_i$, and $x$ may generally be high-dimensional field configurations in a lattice field theory setting. Particular implementations of translationally equivariant convolutional networks are defined in \Cref{sec:translational-sym,sec:group-equiv}. Constructions of coupling layers with affine transformations based on these convolutional networks are defined in \Cref{sec:affine-flows}.}
    \label{fig:coupling-layer-overview}
\end{figure*}

Flow-based models are generally constructed by composing several simple, invertible transformation layers, each described by a number of free parameters. This composition produces an expressive overall transformation that is nevertheless invertible and has a tractable Jacobian determinant. Coupling layers~\cite{dinh2017density} are one common choice of simple transformation in which the degrees of freedom of each sample are divided into two subsets and one subset is updated conditioned on the other, `frozen' subset, as shown in \Cref{fig:coupling-layer-overview}. A `masking pattern' describes the division into subsets. Transformations of the updated subset are parameterized by `context functions' accepting the frozen subset as input, which are typically implemented using neural networks. For example, a simple coupling layer for a real scalar field $\phi(x) \in \mathbb{R}$ could be constructed based on a checkerboard division into even/odd sites, where the field at even sites is (invertibly) transformed by an element-wise rescaling operation plus an additional offset. The scaling factors and offsets are given by the output of an arbitrary context function, which may be parametrized by a neural network acting on the odd sites. The transformation is applied alternatingly between even and odd sites; see Ref.~\cite{Albergo:2021vyo} for a concrete implementation of such coupling layers. Symmetries may be incorporated in such models using appropriate choices of masking patterns, context functions, and transformations. Other choices of layers are also possible (see \Cref{sec:convex-potential-flows} below) and are similarly encoded using generic neural networks.

The target densities defined in \Cref{tab:dists} are all invariant under translations with appropriate boundary conditions, as discussed in \Cref{sec:fermions-sym}. Previous works have shown that exactly incorporating known symmetries into machine learning models can accelerate their training and improve their final quality~\cite{cohen2016group,Cohen:2018,Cohen:2019,Tomiya:2021ywc,Bulusu:2021rqz,Favoni:2020reg}. In the context of normalizing flows, ensuring that the model density is invariant under a symmetry group is achieved by choosing an invariant prior distribution and building transformation layers that are equivariant under the symmetry.  Below, we introduce several `building blocks' which are designed to handle these symmetries. These building blocks are used in the implementation of various layers and flow-based models constructed in this work.

\subsubsection{Translation-equivariant convolutions via P-fields and AP-fields}
\label{sec:translational-sym}

The joint distribution $p(\phi,\varphi)$ given in \Cref{eq:joint-dist} is invariant under simultaneous field translations given by
\begin{equation}
    \phi(\vec{x},t) \rightarrow \phi'(\vec{x},t) = \phi(\vec{x}-\delta\vec{x}, t-\delta t) \label{eq:P-field}
\end{equation}
and
\begin{equation}\label{eq:AP-field}
\begin{aligned}
    \varphi(\vec{x},&t) \rightarrow \varphi'(\vec{x},t) \\
    &= \begin{cases}
      \varphi(\vec{x}-\delta\vec{x}, t-\delta t) & \substack{(t-\delta t) \mathrm{mod} \, 2L_t < L_t} \\
    -\varphi(\vec{x}-\delta\vec{x}, t-\delta t) & \substack{L_t \leq (t - \delta t) \mathrm{mod} \, 2L_t}\ 
    \end{cases} 
\end{aligned}
\end{equation}
for any translations $(\delta \vec{x}, \delta t)$ in the translational symmetry group of the discretized theory. In this work, we label fields transforming as \Cref{eq:P-field} as P-fields, and we label fields transforming as \Cref{eq:AP-field} as AP-fields.\footnote{The fields $\phi$ and $\varphi$ in \Cref{eq:P-field,eq:AP-field}, and P-fields and AP-fields in general, may have multiple components per site.}

In previous applications of flow models to sampling configurations in lattice field theory, translational symmetry has been implemented for bosonic fields by applying convolutional neural networks \cite{lecun-99} with circular padding (periodic boundary conditions) to generate parameters for transformations implemented in each flow layer~\cite{Albergo:2019eim,Nicoli:2020njz,medvidovic2021generative}. All input, intermediate, and output fields in these applications were P-fields. As a building block for translation-equivariant coupling layers acting on both bosonic and pseudofermionic fields, we extend this approach to define translation-equivariant convolutions that act on a generic set of input P-fields and AP-fields, producing output fields with a desired set of transformation properties, i.e., a specification of whether each channel of the output should be a P-field or AP-field. To implement such convolutional neural networks, we exploit the fact that P-fields and AP-fields form an algebra under pointwise addition and multiplication and restrict the operations appropriately to satisfy the desired output transformation properties; see \Cref{app:convnet-fields} for explicit implementation details.

\subsubsection{Translation-equivariant convolutions via group averages}
\label{sec:group-equiv}

As an alternative to defining equivariant convolutional neural networks, one can symmetrize a non-equivariant architecture by explicitly averaging over the whole symmetry group~\cite{cohen2016group}. Convolutional layers with periodic padding in all dimensions are already equivariant under translations of P-fields and under all spatial translations of AP-fields, thus only the subgroup of temporal translations needs to be averaged over to ensure equivariance for AP-fields. The result is a generic method to produce convolutional neural networks with prescribed P-field and AP-field transformation properties of each output channel. Compared with standard convolutions or the restricted equivariant architecture given in \Cref{sec:translational-sym}, this method requires a greater computational effort by a factor proportional to the temporal extent of the lattice, $L_t$. However, it allows the use of unrestricted convolutional architectures, including arbitrary activation functions and learned biases; see \Cref{app:symmetrization} for further details.

\subsubsection{Affine coupling layers}
\label{sec:affine-flows}

\looseness=1
Translation-equivariant networks constructed by either of the methods discussed in \Cref{sec:translational-sym,sec:group-equiv} can immediately be applied in the construction of translation-equivariant affine coupling layers suitable for transforming real-valued scalar fields. An affine coupling layer transforms a field $x$ to $ax + b$ (multiplication and addition are applied pointwise), where $a$ and $b$ are fields produced by context functions acting on the frozen components of the field $x$. Coupling layers, context functions, and masking patterns are illustrated in \Cref{fig:coupling-layer-overview}; see also Ref.~\cite{Albergo:2021vyo}. Using translation-equivariant convolutional neural networks to produce $a$ and $b$, either a bosonic field or pseudofermionic field can be updated in a translation-equivariant manner as long as:

\begin{itemize}
    \item The parameters $a$ and $b$ are both P-fields if $x$ is a bosonic field; or
    \item The parameter $a$ is a P-field and $b$ is an AP-field if $x$ is a pseudofermionic field.
\end{itemize}
Such coupling layers can be composed to produce translation-equivariant flows.

\subsubsection{Equivariant linear operators}
\label{sec:parametric-pf-lin}

The conditional distribution $p(\varphi|\phi)$ is exactly Gaussian, suggesting that it may be efficiently modeled by flows based on architectures other than coupling layers. For example, one may define a linear operator $\mathcal{W}=\mathcal{W}(\phi)$ to transform the pseudofermion fields. The model distribution $q(\varphi| \phi)$ may then be defined by computing ${\varphi = \mathcal{W}\chi}$, where $\chi$ is drawn from the Gaussian distribution $\frac{1}{Z_{\mathcal{N}}}e^{-\chi^\dag \chi}$, such that
\begin{equation}
\label{eq:linear-flow}
\begin{aligned}
    q(\varphi|\phi) &= \frac{1}{Z_\mathcal{N}}e^{-\varphi^\dag (\mathcal{W} \mathcal{W}^\dag)^{-1} \varphi} (\det \mathcal{W} \mathcal{W}^\dag)^{-1} \\
    &= \frac{1}{Z_\mathcal{N}}e^{-\chi^\dag \chi} (\det \mathcal{W} \mathcal{W}^\dag)^{-1}\ .
\end{aligned}
\end{equation}
To effectively use this flow model, $\det(\mathcal{W} \mathcal{W}^\dag)$ must be tractable to compute. In the case of a degenerate pair of fermion flavors, the target distribution is defined by
\begin{equation}
    p(\varphi|\phi) = \frac{1}{Z_{\mathcal{N}} \det{DD^\dag}} e^{-\varphi^\dag [D(\phi) D^\dag(\phi)]^{-1} \varphi}\ .
\end{equation}
While it is clearly sufficient for $\mathcal{W}$ to approximate $D$ in this case, it is in fact only necessary that $\mathcal{W}\mathcal{W}^\dag$ approximates $DD^\dag$, allowing some freedom in the learned matrix~$\mathcal{W}$.

We build the operator $\mathcal{W}$ as a composition of simple linear operators $\mathcal{W}=\mathcal{W}_n\circ\ldots\circ \mathcal{W}_1$, where each $\mathcal{W}_k$ has only local interactions along a fixed dimension, in a fixed direction (that is, with only positive or negative offsets, but not both), allowing the determinant of each matrix to be efficiently computed. We choose to parameterize the components of each operator $\mathcal{W}_k$ by two P-fields, produced from learned translation-equivariant functions of $\phi$. More specifically, we consider $2d$ types of operators, where each type is defined by a sign $s=\pm 1$ and a choice of one of the $d$ lattice directions. For the two-dimensional application described below, there are thus four distinct operator types. The different types of operators are applied alternatingly in the composition, but the specific order can be chosen arbitrarily. The operator type with couplings in the spatial direction and sign $s$ thus updates a field $\chi$ by
\begin{equation}
    (\mathcal{W} \, \chi)_{ij} = a_{ij}\chi_{ij} + b_{ij}\chi_{i+s,j}
\end{equation}
with periodic boundary conditions along the space dimension: $\chi_{L+1,j}=\chi_{1, j}$ and $\chi_{0, j}=\chi_{L, j}$. An operator with temporal couplings updates a field $\chi$ by
\begin{equation}
    (\mathcal{W} \, \chi)_{ij} = a_{ij}\chi_{ij} + b_{ij}\chi_{i,j+s}
\end{equation}
with antiperiodic boundary conditions along the time dimension: $\chi_{i, L+1}=-\chi_{i, 1}$ and $\chi_{i, 0}=-\chi_{i, L}$. This construction may be understood as a convolutional layer with appropriate boundary conditions and an additional constraint on the kernel to have non-zero entries only in the center and at one of the $2d$ adjacent sites.

With these definitions, each operator $\mathcal{W}_k$ is block diagonal (for a suitable choice of basis). Each block is of the form
\begin{equation}
\begin{aligned}
    \begin{bmatrix}
    a_1 &  &  &  & \pm b_1\\ 
    b_2 &  a_2 &  & 0  & \\ 
     & \ldots & \ldots &  & \\ 
     & 0 &  &  & \\ 
     &  &  & b_L & a_L
    \end{bmatrix}\ ,
\end{aligned}
\end{equation}
where we have dropped a (spatial or time) index to simplify the notation. The determinant of each block is simply $\Pi_h a_h \pm \Pi_h b_h$, indicating that the Jacobian determinant associated with the full composition can be tractably computed.

\subsubsection{Convex Potential Flows}
\label{sec:convex-potential-flows}

Because of the non-local nature of the effective action, we consider an alternative flow architecture to produce a model distribution $q(\phi)$ approximating $p(\phi)$. Convex Potential Flows (CPFs) are normalizing flows defined via the gradients of a potential that is strongly convex and twice differentiable almost everywhere~\cite{ZhangEWang2018Monge, huang2020convex}. Strong convexity of the potential on a convex support $\mathcal{X}$ guarantees the flow to be invertible on $\mathcal{X}$, and this family of normalizing flows can be shown to be a universal density approximator~\cite{huang2020convex}. We can parameterize strongly convex functions by neural networks with mild constraints on their architecture and weights~\cite{pmlr-v70-amos17b}.

More specifically, given a convex potential function ${u: \mathcal{X} \rightarrow \mathbb{R}}$, we define the map
\begin{equation}\label{eq:CPF}
    [f(z)]_i = \frac{\partial}{\partial z_i} u(z)\ ,
\end{equation}
where the index $i$ specifies how each degree of freedom of $z$ is mapped. Starting from a base density $r(z)$, the resulting probability density produced by mapping through $f$ follows from \Cref{eq:logq-generic} as
\begin{equation}
    q(x) = r(z)\det H_u(z) ^{-1}\ ,
\end{equation}
where $x = f(z)$ and $H_u(z)=\frac{\partial^2}{\partial z_i \partial z_j} u(z)$ is the Hessian matrix of $u(z)$. Training by minimizing the KL-divergence $D_{\KL}(q||p)$ between the model $q$ and a target density $p$ only requires the gradients $\nabla_{\theta} \log \det H_u(x)$ with respect to the model's parameters $\theta$. Since the Hessian is symmetric and positive-definite for strongly convex potentials, we can directly employ a stochastic trace estimator~\cite{huang2020convex,dong2017scalable},
\begin{equation}\label{eq:grad.log.det.H}
\begin{aligned}
    \nabla \log \det H_u(x) &= \nabla \Tr \log H_u(x) \\
    &= \Tr\left[ H_u(x)^{-1} \nabla H_u(x)\right] \\
    &=\mathbb{E}_{\chi \sim e^{-\chi^\dag \chi}}[(H^{-1}_u(x) \chi)^\dag \nabla H_u(x)\chi]\ .
\end{aligned}
\end{equation}
The sample mean over noise vectors $\chi$ can be used to estimate this quantity in practice, and the inverse Hessian applied in $H_u(x)^{-1}\chi$ can be efficiently computed by the application of the conjugate-gradient method. Note that this estimator only requires the computation of Hessian-vector products $H \chi$, which is particularly convenient when the Hessian is sparse.

CPFs can be straightforwardly applied to construct flows that sample bosonic fields $\phi$. They can also be constrained to be translation-equivariant by using appropriate convolutions. In contrast to coupling layers, the CPF potential is a scalar function based on global information, which may result in transformations of the field $\phi$ that can in general be quite non-local. Evaluating the model probability density for use in asymptotically exact Markov chains requires a precise approximation of the log-det Hessian to avoid systematic errors. An exact calculation of the determinant is feasible only for small lattice volumes. For larger field configurations, one could apply a more scalable estimator, such as the estimator based on Lanczos tridiagonalization and the quadrature method described in Ref.~\cite{ubaru2017fast}.

\subsection{Flow models}
\label{sec:flow-models}

We next define particular architectures for modeling each of the distributions required for the four sampling approaches introduced in \Cref{sec:sampling-dists}. While the space of possible architectures that may be defined from the building blocks of \Cref{sec:building-blocks} is large and the present discussion is not exhaustive, the use of each sampling method and each building block is demonstrated at least once. The architectures for each approach detailed in this section are summarized in \Cref{fig:models-arch}.

\subsubsection{Modeling \texorpdfstring{$p(\phi)$}{p(phi)} for \texorpdfstring{$\phi$-marginal}{phi-marginal} sampling}
\label{sec:phi-marginal-modeling}

The $\phi$-marginal sampler defined in \Cref{sec:phi-marginal-sampling} requires a flow whose model distribution $q(\phi)$ approximates $p(\phi)$. Such a flow only needs to manipulate P-fields. We build this $\phi$-marginal model using a composition of CPF layers, where the output of each layer is defined by computing the gradient of a potential $u_i(\cdot)$ (see \Cref{sec:convex-potential-flows}). These layers act on samples $\zeta$ drawn from some base distribution $r_{\mathrm{p}}(\zeta)$. \Cref{fig:marginal-phi-arch} depicts this type of $\phi$-marginal architecture defined by a composition of CPFs acting on $\zeta$. Each $u_i$ contributes a determinant factor $\det H_{u_i}^{-1}$ to $q(\phi)$ in terms of the Hessian $H_{u_i}^{ab} = \frac{\partial^2 u_i(z)}{\partial z_a \partial z_b}$, such that
\begin{equation}
    q(\phi) = r_{\mathrm{p}}(\zeta)\prod_i \det H_{u_i} ^{-1}\ .
\end{equation}
As discussed in \Cref{sec:flow-optim}, this marginal model is extended to the joint density $q(\phi, \varphi) = q(\phi)p(\varphi | \phi)$ for training. The density cannot be computed efficiently due to the determinants involved in the definition of $q(\phi)$ as well as in the normalizing constant of $p(\varphi|\phi)$, but the flow is nevertheless trainable using stochastic estimates of the gradients. For sampling, the joint density itself may also be estimated using stochastic approximations of the determinant factors.

\looseness=1
The architecture of the convex potential network $u(\phi)$ is based on Ref.~\cite{huang2020convex} and is modified appropriately to account for the periodic boundary conditions. It consists of $K$ layers of convolutions of the form
\begin{equation}\label{eq:cpf-conv}
\begin{aligned}
    h_1 &= L_1(\phi) \\
    h_{k+1} &= L_k^+(\text{SoftPlus}(\text{ActNorm}(h_k))) + L_k(\phi) \\
    u(\phi) &= w_1 \text{Sum}(h_K) + w_2 \frac{\| \phi \|^2}{2}\ ,
\end{aligned}
\end{equation}
where $L_j$ is a convolution layer with periodic boundary conditions and unconstrained weights; $L_j^+$ is a convolution layer with periodic boundary conditions and positive-only weights; $\text{SoftPlus}(x) = \log(1 + \exp(x))$; $\text{ActNorm}(x)=(x - \mu)/\sigma$ is layer that normalizes its inputs using a learnable offset $\mu$ and scale $\sigma$, where $\mu$ and $\sigma$ are initialized as the mean and standard deviation of the inputs of an initialization batch~\cite{NEURIPS2018_d139db6a}; $w_1,w_2$ are learnable weights used to control closeness of the flow to the identity map at initialization. The use of periodic boundary conditions for $L_j$ and $L_j^+$ and the final Sum operation ensures that $u(\phi)$ is invariant to translations.

\subsubsection{Modeling \texorpdfstring{$p(\phi|\varphi)$}{p(phi|varphi)} for Gibbs sampling}
\label{sec:gibbs-modeling}

\begin{figure*}
    \centering
    \setlength{\fboxrule}{0pt}
    
    \subfloat[$\phi$-Marginal architecture based on convex potential flows (\Cref{sec:phi-marginal-modeling}).]{
    \setlength{\fboxsep}{0.7cm}
    \fbox{%
    \includegraphics{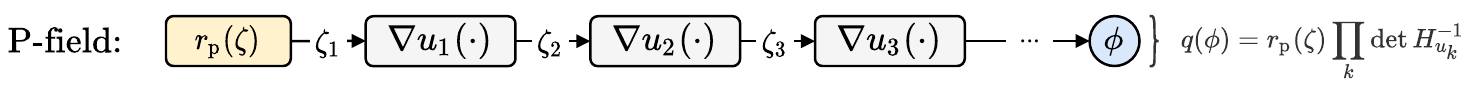}\label{fig:marginal-phi-arch}
    }
    }
    
    \subfloat[Fully joint architecture for $q(\phi,\varphi)$ based on coupling layers (\Cref{sec:joint-modeling}).]{
    \setlength{\fboxsep}{0.5cm}
    \fbox{%
    \includegraphics{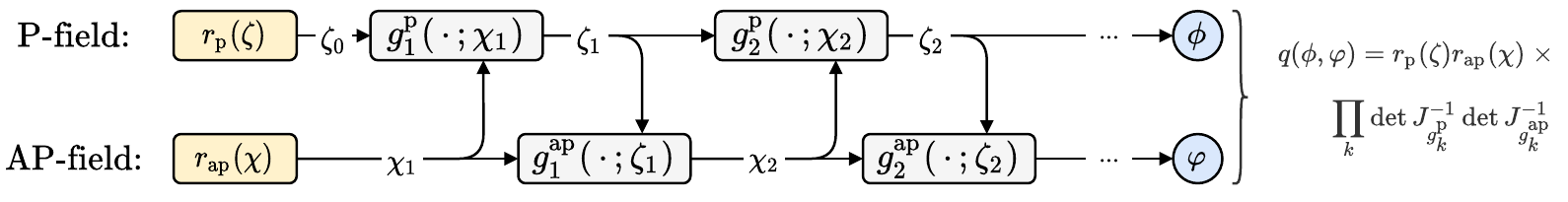}\label{fig:fully-joint-arch}
    }
    }
    
    \subfloat[$\phi$-Conditional model $q(\phi|\varphi)$ defined via a restricted joint architecture (\Cref{sec:gibbs-modeling}).]{
    \setlength{\fboxsep}{0.5cm}
    \fbox{%
    \includegraphics{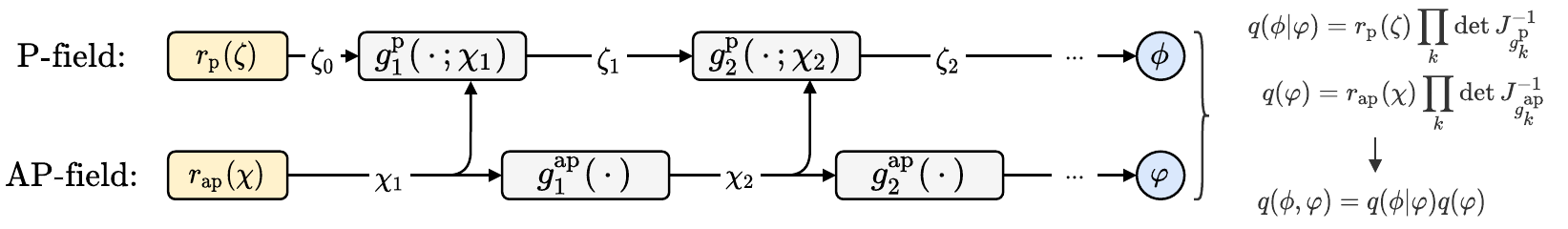}\label{fig:phi-cond-varphi-arch}
    }
    }
    
    \subfloat[Autoregressive model $q(\phi)q(\varphi|\phi)$ defined via coupling layers and linear flows (\Cref{sec:autoregressive-modeling}).]{
    \setlength{\fboxsep}{0.5cm}
    \fbox{%
    \includegraphics{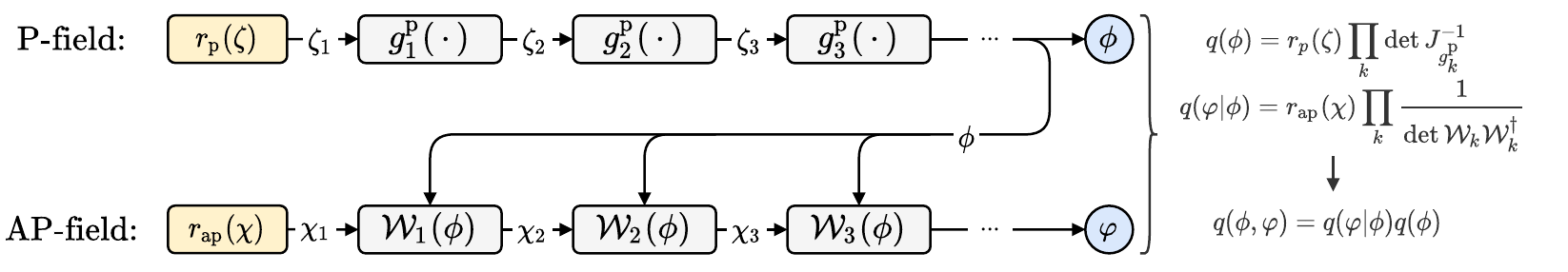}\label{fig:varphi-linear-arch}
    }
    }
    \caption{Architectures for the flow-based models defined in \Cref{sec:flow-models} for each sampling approach.
    %Components highlighted in red for the conditional and joint models reflect the restriction applied to move from a fully joint to conditional architecture.
    Note that each coupling layer $g^{\mathrm{p}}_k$ or $g^{\mathrm{ap}}_k$ employs masking of the updated field as shown in \Cref{fig:coupling-layer-overview}, such that the frozen components of the field are included as input to context functions. Superscripts on coupling layers indicate the translational equivariance structure of coupling layer inputs and outputs (either consistently transforming as P-fields or AP-fields). Many other choices of architectures are possible to model each distribution; the figure reflects the choices utilized in the numerical study undertaken in \Cref{sec:yukawa-theory}.}
    \label{fig:models-arch}
\end{figure*}

\Cref{sec:gibbs-sampling} describes a Gibbs sampling scheme that utilizes the exact conditional $p(\varphi|\phi)$ and a modeled conditional density $q(\phi|\varphi)$. A $\varphi$-marginal model $q(\varphi)$ is required to extend $q(\phi|\varphi)$ to the joint distribution $q(\phi|\varphi)q(\varphi)$ for training. We achieve this simultaneous modeling of $q(\phi|\varphi)$ and $q(\varphi)$ by using a fully joint architecture with restricted information flow, as shown in \Cref{fig:phi-cond-varphi-arch}. The model consists of a prior distribution over the base configurations $\zeta, \chi$ denoted by $r_{\mathrm{p}}(\zeta)$ and $r_{\mathrm{ap}}(\chi)$, followed by the application of two types of affine coupling layers. First, the layers $g^{\mathrm{p}}_{k}(\cdot; \chi_k)$ update the P-field configuration conditioned on the AP-field, along with the frozen components of $\zeta_k$, to produce $q(\phi | \varphi)$ as: 
\begin{equation}
    q(\phi | \varphi) = r_{\mathrm{p}}(\zeta) \prod_{k} \det J_{g^{\mathrm{p}}_k}^{-1},
\end{equation}
where $J_{g^{\mathrm{p}}_k}$ is the Jacobian for coupling $g^{\mathrm{p}}_k$.
Second, the couplings $g^{\mathrm{ap}}_{k}(\cdot)$ transform the AP-field $\chi$ conditioned solely on its frozen components to obtain $q(\varphi)$:
\begin{equation}
    q(\varphi) = r_{\mathrm{ap}}(\chi) \prod_{k} \det J_{g^{\mathrm{ap}}_k}^{-1}
\end{equation}

To conditionally re-sample $\phi$ from $q(\phi|\varphi)$ while leaving $\varphi$ unchanged, the bosonic prior variable is re-sampled and the output of the flow is re-evaluated while holding the pseudofermionic prior variable $\chi$ fixed. When $\varphi$ is re-sampled from $p(\varphi|\phi)$ in the alternate step of the Gibbs sampler, it is important to update the value of $\chi$ by passing $\varphi$ through the inverse of the bottom branch of the flow depicted in the figure. This allows future re-sampling of $\phi$ as well as the calculation of the conditional probability density defined by the model.

\subsubsection{Autoregressive modeling of \texorpdfstring{$p(\phi,\varphi)=p(\phi)p(\varphi|\phi)$}{p(phi,varphi)=p(phi)p(varphi|phi)}}
\label{sec:autoregressive-modeling}

\Cref{sec:autoregressive-sampling} defines an independence sampler based on an autoregressive joint model with the probability density given by $q(\phi, \varphi) = q(\phi)q(\varphi | \phi)$. We implement $q(\phi)$ using masked affine coupling layers, as was done in Refs.~\cite{Albergo:2019eim,Kanwar:2020xzo,Boyda:2020hsi}. The parameters of the affine coupling layers are given by convolutional networks satisfying translational equivariance through standard periodic boundaries. We implement $q(\varphi|\phi)$ using a deep linear flow consisting of learned linear operators $\mathcal{W}_k(\phi)$ as detailed in \Cref{sec:parametric-pf-lin}. The parameters of these linear operators are all P-fields obtained by similar periodic convolutional networks. The full joint model is given by the autoregressive combination of these two models, i.e.~drawing $\phi$ from the affine model with distribution $q(\phi)$, then drawing $\varphi$ from the conditional deep linear flow with distribution $q(\varphi|\phi)$, as shown in \Cref{fig:varphi-linear-arch}. The marginal model is defined by sampling $\zeta$ from the prior distribution $r_{\mathrm{p}}(\zeta)$, then applying the sequence of coupling layers $g^{\mathrm{p}}_k(\cdot)$ such that the marginal model probability density $q(\phi)$ is given by:
\begin{equation}
    q(\phi) = r_{\mathrm{p}}(\zeta)\prod_k \det J_{g^{\mathrm{p}}_k}^{-1}\ .
\end{equation}
The conditional linear flow is defined by sampling $\chi$ from the prior distribution $r_{\mathrm{ap}}(\chi)$ and applying the linear operators $\mathcal{W}_k(\phi)$ to obtain the model density
\begin{equation}
    q(\varphi | \phi) = r_{\mathrm{ap}}(\chi) \prod_k \frac{1}{\det \mathcal{W}_k \mathcal{W}^\dagger_k}\ .
\end{equation}
We define $r_{\mathrm{ap}}(\chi) = \frac{1}{Z_{\mathcal N}} e^{-\chi^\dagger\chi}$ to match the choice for the linear operator flow in \Cref{eq:linear-flow}.

Note that the learned components in this approach may also be combined in novel ways. For example, it is possible to discard the conditional flow with distribution $q(\varphi|\phi)$ after training and simply use $q(\phi)$ for $\phi$-marginal sampling as described in \Cref{sec:phi-marginal-sampling,sec:phi-marginal-modeling}. This may be advantageous in situations where gradients from an exactly sampleable distribution are not available and training must be fully variational. On the other hand, the conditional deep linear flow may be used by itself as a determinant estimator for given configurations $\phi$.

\subsubsection{Fully joint modeling of \texorpdfstring{$p(\phi,\varphi)$}{p(phi,varphi)}}
\label{sec:joint-modeling}

Finally, we construct a model that simultaneously samples $\phi$ and $\varphi$ in a fully joint approach which can be employed in the exact sampler defined in \Cref{sec:joint-sampling}. The joint model implemented in this work is constructed from affine coupling layers that alternatingly transform the bosonic fields conditioned on the pseudofermionic fields, and vice versa, as shown in \Cref{fig:fully-joint-arch}. The model is defined to sample $\zeta$ and $\chi$ from the prior distributions $r_{\mathrm{p}}(\zeta)$, $r_{\mathrm{ap}}(\chi)$ and subsequently apply alternating layers. Coupling layers $g^{\mathrm{p}}_{k}(\cdot, \chi_k)$ transform the P-field base configuration $\zeta$ conditioned on its frozen components and the AP-field configuration $\chi_k$, while couplings $g^{\mathrm{ap}}_{k}(\cdot, \zeta_k)$ update the AP-field base configuration $\chi_k$ conditioned on its frozen components and $\zeta_k$. This gives rise to the joint density
\begin{equation}
    q(\phi, \varphi) = r_{\mathrm{p}}(\zeta) r_{\mathrm{ap}}(\chi) \prod_k \det J_{g^{\mathrm{p}}_k}^{-1} \det J_{g^{\mathrm{ap}}_k}^{-1}\ .
\end{equation}

\section{Application to a scalar Yukawa theory in two dimensions}
\label{sec:results}

As a demonstration, we implement the sampling algorithms defined above for a two-dimensional model of a scalar field coupled to fermions via a Yukawa interaction. This model provides a testbed which features fermionic fields, but without the additional complications brought on by gauge symmetry. Apart from providing a suitable test case for the development of flows capable of modeling fermions, studying Yukawa interactions is also interesting in its own right, e.g.~for Higgs physics~\cite{Witzel:2019jbe} or the quark-meson model~\cite{Wambach:2009ee}.

\subsection{Yukawa theory on the lattice}
\label{sec:yukawa-theory}

We consider a real, scalar field $\phi$ coupled to one mass-degenerate pair of Kogut-Susskind staggered fermions~\cite{Kogut:1974ag}. We emphasize that our method can in principle be straightforwardly applied to other discretization schemes, such as Wilson fermions, without additional conceptual difficulties. 

The action for this theory is defined as $S(\psi,\bar{\psi},\phi) = S_B(\phi) + S_F(\psi, \bar{\psi}, \phi)$ (see \Cref{eq:fermionic-theory-action}). For the scalar field we choose the usual discretized Klein-Gordon action with quartic self-interaction, defined by
\begin{equation}
\begin{aligned}
    S_B(\phi) = \sum_{x\in\Lambda}\Bigr[-2 & \sum_{\mu=1}^{d} \phi(x)\phi(x+\hat{\mu}) \\
    &+ (m^2 + 2d)\phi(x)^2 + \lambda\phi(x)^4\Bigr]\ ,
\end{aligned}
\end{equation}
where $\Lambda$ denotes the set of lattice sites, $m$ the bare scalar mass parameter, $\lambda$ the coupling, and $d$ the dimension. The fermionic action $S_F$ is defined by the bilinear form in \Cref{eq:fermion-action} with $N_f = 2$ and both fermion flavors are defined by the discretized Dirac operator
\begin{equation}\label{eq:dirac-op}
\begin{aligned}
    D_{xy} = \sum_{\mu = 1}^{d} & \eta_\mu(x) \frac{\delta(x-y+\hat{\mu})-\delta(x-y-\hat{\mu})}{2}  \\
    &+ \delta(x - y)(m_f + g\phi(x))\ ,
\end{aligned}
\end{equation}
where $m_f$ is the bare mass of the fermion and $g$ the Yukawa coupling. The staggered factor $\eta_\mu$ is obtained from the Dirac $\gamma$-matrices after the staggered transformation and is defined as
\begin{equation}
    \eta_1(x)=1 \quad \textrm{and} \quad \eta_l(x) = (-1)^{x_1}\cdots(-1)^{x_{l-1}}\ .
\end{equation}
The Kronecker $\delta$ is defined to have antiperiodic boundary conditions in the time direction (conventionally taken to be $\mu = d$) and periodic boundary conditions in the spatial directions, i.e.
\begin{equation}
    \delta (x) = \prod_{\mu = 1}^d \delta_\mu (x_\mu)
\end{equation}
where
\begin{equation}
\begin{aligned}
  \delta_{\mu \neq d}(x_{\mu}) &=
  \begin{cases}
    1 & \text{if $x_\mu = 0, \pm L$} \\
    0 & \text{otherwise}
  \end{cases} \\
  \text{and} \quad
  \delta_d(x_d) &=
  \begin{cases}
    1 & \text{if $x_d = 0$} \\
    -1 & \text{if $x_d = \pm L_t$} \\
    0 & \text{otherwise.}
  \end{cases}
\end{aligned}
\end{equation}

We employ an even-odd preconditioning scheme for the Dirac operator for all models except for the autoregressive model using linear operators. In contrast to the default lexicographic ordering, sorting lattice sites into even and odd allows to bring the matrix into a form that is amenable to an explicit block factorization of the determinant, which leads to improvements in the conditioning and solver performance. This reduces the variance and cost of computing the pseudofermion action required for optimizing models and sampling. Most previous work on improved orderings has focused on techniques for Wilson fermions in the context of gauge theory~\cite{Frommer:1994vn,deForcrand:1996ck,Peardon:2000si}, but the same insights can be applied to the staggered fermion formulation used in this work, as detailed in \Cref{app:even-odd}.

All results reported in this work are computed on a $16 \times 16$ lattice geometry using the two choices of action parameters in the symmetric phase given in \Cref{tab:params}. For this theory and lattice discretization, there is no additive renormalization to the bare fermion mass $m_f$. Accordingly, we directly probe the case of vanishing mass by setting $m_f = 0$. The first set of parameters, for which the Yukawa coupling is chosen to be $g=0.1$, already provides a realistic test scenario in the sense that the average ratio of fermionic to scalar force magnitudes is around $3\%$, which is similar to the ratio of fermionic to gauge forces reported in the literature for some lattice QCD computations; see e.g.~Refs.~\cite{Luscher:2004pav,Urbach:2005ji}. The second choice with $g=0.3$ features a much larger force ratio amounting to about $39\%$, which thus provides a testbed for theories with more prominent fermionic effects. For simplicity, we will refer to these two parameter choices by the associated value of the Yukawa coupling $g$.

\begin{table}[t]
    \begin{ruledtabular}
    \begin{tabular}{ccccc >{\arraybackslash\hspace{1ex}}ccc}
    \centering
    $V$ & $m^2$  & $\lambda$ & $g$ & $m_f$ & $\langle|M|\rangle$ & $\langle|\bar\psi \psi|\rangle$ & Force ratio \\ \midrule
    $16^2$ & $-4.00$ & $6.0$ & $0.1$ & $0$ & $0.0733(1)$ & $0.0159(1)$ & $3\%$ \\
    $16^2$ & $-1.55$ & $2.4$ & $0.3$ & $0$ & $0.0791(1)$ & $0.0490(1)$ & $39\%$ \\
    \end{tabular}
    \end{ruledtabular}
    \caption{The two parameter choices for our reported numerical studies and the associated average absolute magnetization and chiral condensate computed with HMC. All uncertainties reported in this work are obtained using data blocking to account for autocorrelations and applying the statistical jackknife method. Force ratios are determined by dividing the average $L^2$-norms of fermionic and bosonic force vectors.}
    \label{tab:params}
\end{table}

To evaluate the performance of the sampling approaches presented here, we consider the following quantities and observables. The magnetization of a scalar field configuration is defined as
\begin{equation}
    M = \frac{1}{V} \sum_{x \in \Lambda} \phi(x)\ .
\end{equation}
We measure the average absolute value $\langle|M|\rangle$, which provides a non-zero order parameter that is large in the broken symmetry phase and exponentially suppressed in the symmetric phase. The connected two-point correlation function of $\phi$ is defined as
\begin{equation}
    C_\phi(x,y) = \langle \phi(x) \phi(y) \rangle - \langle \phi(x) \rangle \langle \phi(y) \rangle \ ,
\end{equation}
where we fix $\langle \phi(x) \rangle = \langle M \rangle = 0$ analytically. We evaluate the source-averaged correlator projected to zero momentum defined by
\begin{equation}
    C_\phi(t) = \frac{1}{V} \sum_{x} \sum_{\vec{y}} C(x, x+(\vec{y},t))\ .
\end{equation}
Fermionic observables can be computed from the matrix elements of the inverse Dirac operator. The chiral condensate of the fermion field is defined by
\begin{equation}
    \langle \bar\psi\psi \rangle = \left\langle \frac{1}{V} \mathrm{Tr}\, D^{-1} \right\rangle \ ,
\end{equation}
and we measure $\langle|\bar\psi\psi|\rangle$ as for the magnetization. Using the off-diagonal matrix elements, we also evaluate the average fermionic two-point correlator in the time direction,
\begin{equation}
    C_\psi(t) = \left< \psi(y) \,  \bar{\psi}(0) \right> = \left< D^{-1}_{y,0} \right> \ ,
\end{equation}
where $y = (\vec{0}, t)$ with $t$ odd. The particular choices of offsets $y$ select staggered spinor components at the sink $\psi(y)$ that result in a non-zero average correlation function originating from the source $\bar{\psi}(0)$.

\begin{table*}
    \centering
    \begin{ruledtabular}
    \begin{tabular}{m{1.3in}>{\centering\arraybackslash}p{1.0in}ccccccc}
         MCMC Approach & Modeled targets & Flow model & Acc.\ rate & $\langle|M|\rangle$ & $\langle|\bar\psi \psi|\rangle$ & $\tau^{\text{int}}_{M}$ & $\tau^{\text{int}}_{\bar{\psi} \psi}$ \\
    \midrule
         $\phi$-Marginal (\ref{sec:phi-marginal-sampling}) & $p(\phi)$ & \ref{sec:phi-marginal-modeling} & $92\%$ & $0.0734(1)$ & $0.0159(1)$ & $0.72(1)$ & $0.71(1)$ \\
         & & & $92\%$ & $0.0792(1)$ & $0.0491(1)$ & $0.67(1)$ & $0.67(1)$ \\ \vspace{1ex}
         Gibbs (\ref{sec:gibbs-sampling}) & $p(\phi|\varphi)$ & \ref{sec:gibbs-modeling} & $60\%$ & $0.0735(1)$ & $0.0160(1)$ & $2.02(4)$ & $2.02(3)$ \\
         & & & $44\%$ & $0.0792(1)$ & $0.0490(1)$ & $2.74(4)$ & $2.73(4)$ \\ \vspace{1ex}
         Autoregressive (\ref{sec:autoregressive-sampling}) & $p(\phi), p(\varphi|\phi)$ & \ref{sec:autoregressive-modeling} & $53\%$ & $0.0731(1)$ & $0.0159(1)$ & $2.16(3)$ & $2.16(3)$  \\
         & & & $43\%$ & $0.0790(1)$ & $0.0489(1)$ & $3.62(7)$ & $3.60(7)$ \\ \vspace{1ex}
         Fully Joint (\ref{sec:joint-sampling}) & $p(\phi,\varphi)$ & \ref{sec:joint-modeling} & $37\%$ & $0.0733(1)$ & $0.0159(1)$ & $4.98(11)$ & $4.98(11)$ \\
         & & & $31\%$ & $0.0791(1)$ & $0.0490(1)$ & $8.73(30)$ & $8.67(30)$ \\
    \end{tabular}
    \end{ruledtabular}
    \caption{Sampling performance metrics and observables for all approaches, computed from 100 Markov chains with 10k proposals each, where the first 1k are discarded for thermalization. For each model, the first row shows results obtained for $g = 0.1$ and the second row for $g = 0.3$, respectively. For comparison, the values obtained with HMC listed in \Cref{tab:params} are consistent with the measurements from our models. Autocorrelation times $\tau^{\text{int}}$ are computed for each of the 100 chains and then averaged, and errors are obtained with statistical jackknife. The results are discussed in more detail in \Cref{sec:results-discussion}. All models except the autoregressive make use of even-odd preconditioning of the action.}
    \label{tab:model-metrics}
\end{table*}

\subsection{Model architectures}
\label{sec:model-archs}

For each of the four sampling approaches outlined in \Cref{sec:sampling-dists} and corresponding model architectures detailed in \Cref{sec:flow-models}, we create specific models for both choices of target action parameters given in \Cref{tab:params}:
\begin{itemize}
    \setlength\itemsep{1ex}
    \item For the sampling scheme described in \Cref{sec:phi-marginal-sampling}, we construct a CPF model defining a $\phi$-marginal distribution $q(\phi)$ approximating the corresponding $p(\phi)$. The model architecture and training follow the generic procedure outlined in \Cref{sec:phi-marginal-modeling};
    \item To build a conditional model $q(\phi|\varphi)$ for the Gibbs sampler described in \Cref{sec:gibbs-sampling}, we implement a restricted affine coupling layer flow as described in \Cref{sec:gibbs-modeling}. To achieve translational equivariance, we employ the method of group averages described in \Cref{sec:group-equiv} for each convolutional network;
    \item To produce an autoregressive joint model density $q(\phi,\varphi) = q(\phi)q(\varphi|\phi)$ for the sampling scheme described in \Cref{sec:autoregressive-sampling}, we use a model consisting of affine coupling layers followed by learned equivariant linear transformations as described in \Cref{sec:autoregressive-modeling};
    \item For the fully joint sampling scheme described in \Cref{sec:joint-sampling}, we implement a model with unrestricted affine coupling layers acting on both $\phi$ and $\varphi$, using translation-equivariant convolutions as described in \Cref{sec:translational-sym}. This results in a fully joint model distribution $q(\phi,\varphi)$ as detailed in \Cref{sec:joint-modeling}.
\end{itemize}
The models are optimized for each approach based on the joint KL divergence discussed in \Cref{sec:flow-optim}. Prior distributions for the initial P-field $\zeta$ and AP-field $\chi$, where they are used according to \Cref{fig:models-arch}, are Gaussians of the form
\begin{equation}
\begin{aligned}
    r_{\mathrm{p}}(\zeta) &= \frac{1}{\mathcal {Z_{\zeta}}} e^{-\zeta^{\dagger} \zeta / (\sigma_{\zeta})^2 } \\
    \mathrm{and}\ r_{\mathrm{ap}}(\chi) &= \frac{1}{\mathcal {Z_{\chi}}} e^{-\chi^{\dagger} \chi / (\sigma_{\chi})^2}\ ,
\end{aligned}
\end{equation}
with specific values of $\sigma_{\zeta}$ and $\sigma_{\chi}$ for each model chosen to enhance the training stability.

Details of the model hyperparameters and training procedure for each of the the models can be found in \Cref{app:details}. An exhaustive search over the available parameter space is beyond the scope of the present work, and as such it can be expected that tuning the various model hyperparameters may further improve the reported performance metrics.

\subsection{Discussion and comparison of sampling schemes}
\label{sec:results-discussion}

After optimization, we use each of the models to construct asymptotically exact samplers for their respective target distributions according to the four schemes given in \Cref{sec:sampling-dists}. For each case, we produce 100 distinct Markov chains consisting of 10k steps each, of which the first 1k steps are discarded for thermalization. These Markov chains are used for observable measurements and to investigate and compare metrics of the efficiency of sampling via each of these methods.

\begin{figure*}
    \centering
    \includegraphics{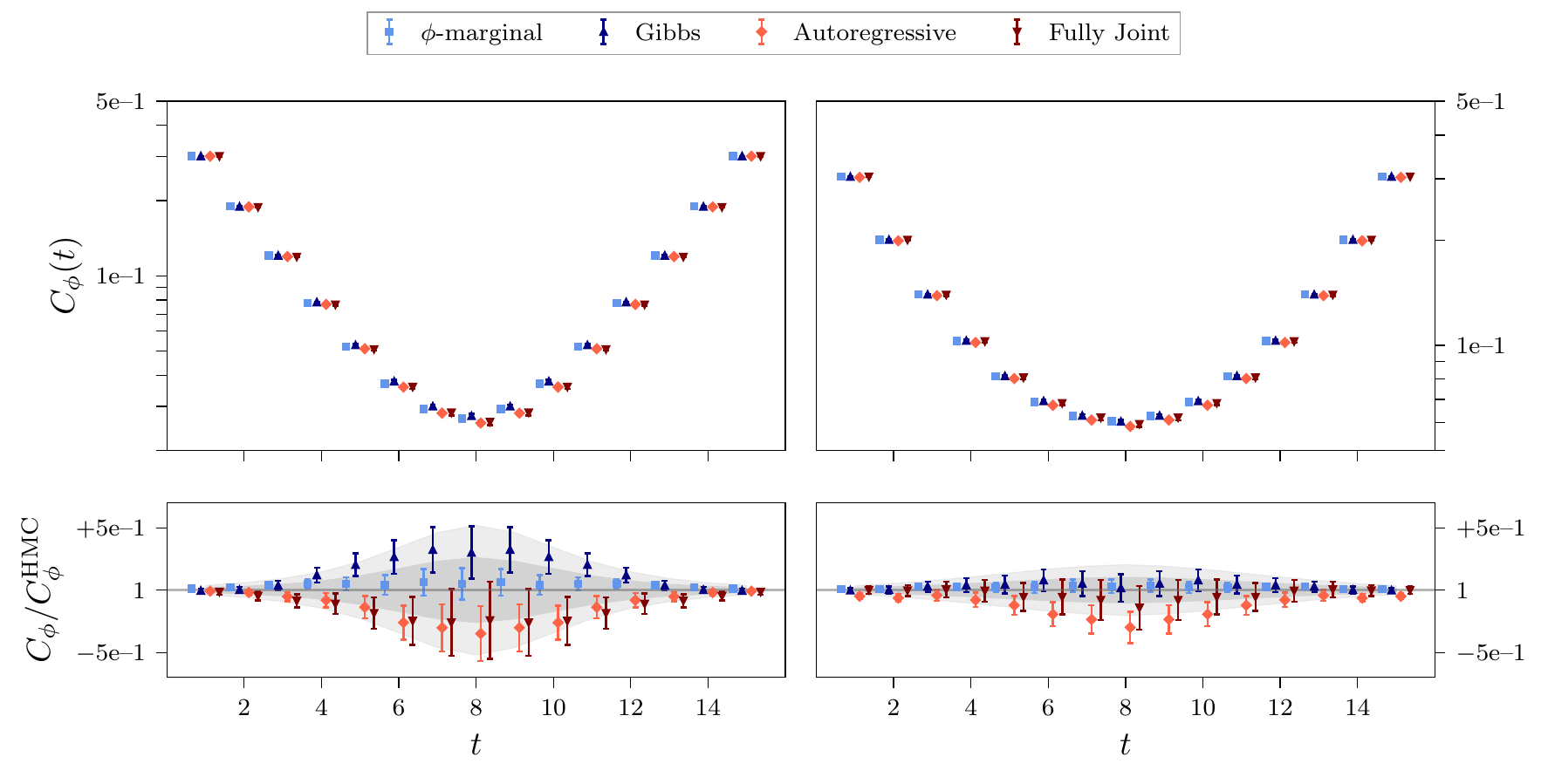}
    \caption{Connected two-point correlation functions of the scalar field (projected to zero spatial momentum in each time slice $t$) for each model and choice of action parameters, computed from 100 Markov chains with 9k configurations each. Error estimates are obtained using data blocking with a bin size of 100 and applying statistical jackknife. Left: $g = 0.1$, right: $g = 0.3$. Bottom panels show the ratio of each data point to the HMC baseline, where the shaded regions correspond to the $1\sigma$ and $2\sigma$ uncertainty bands of the HMC results. For both the scalar correlators here and the fermionic ones in \Cref{fig:fermion-correlators}, Hotelling's t-squared statistic~\cite{hotelling} comparing each flow model result to the HMC baseline finds results to be consistent with correlated statistical fluctuations.}
    \label{fig:scalar-correlators}
\end{figure*}

\begin{figure*}
    \centering
    \includegraphics{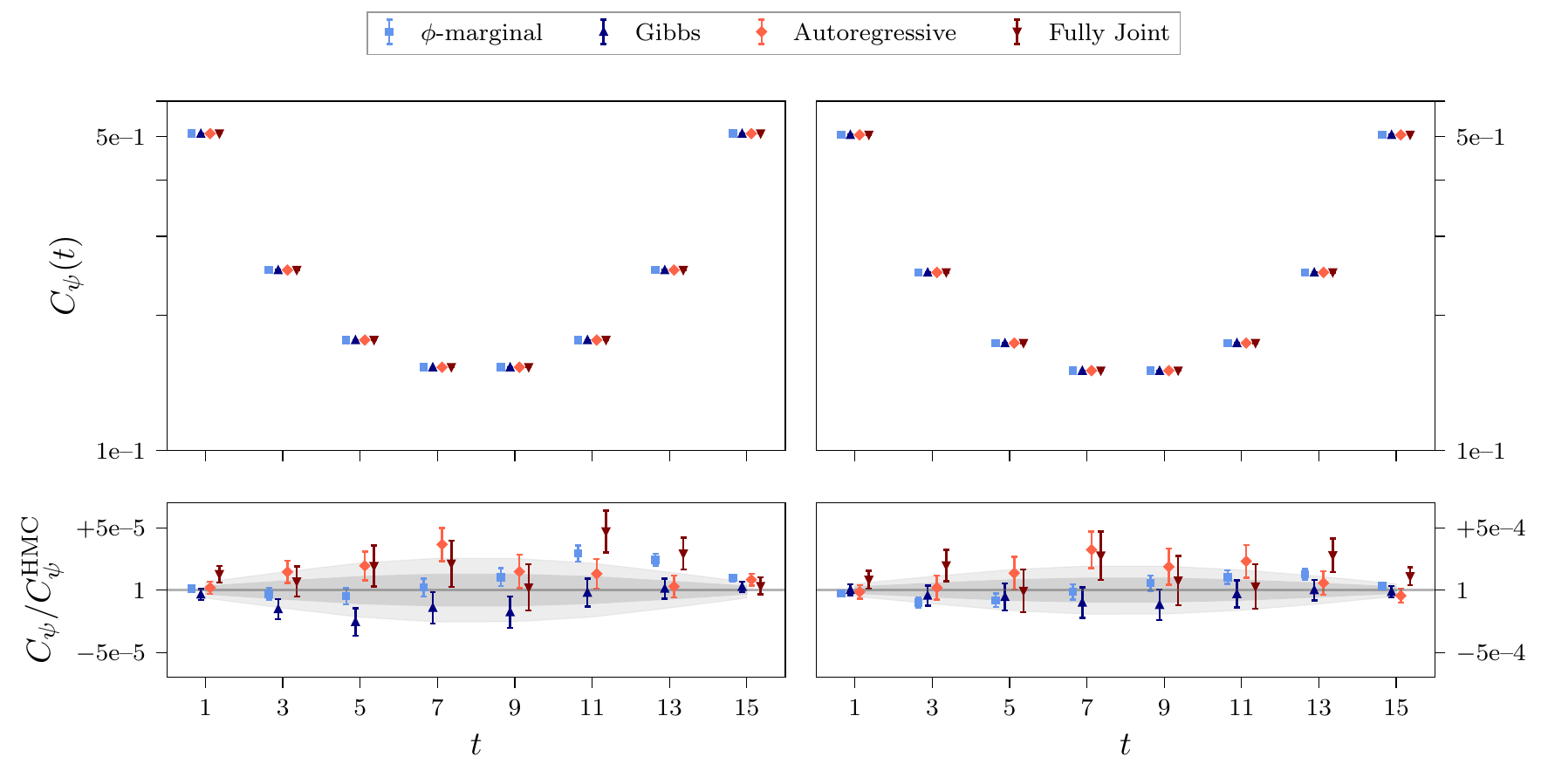}
    \caption{Average fermionic two-point correlation in the time direction for each model and choice of action parameters using the same configurations and data blocking as for \Cref{fig:scalar-correlators}. Left: $g = 0.1$, right: $g = 0.3$. The choice of odd $t$ selects staggered spinor components at the sinks that give a non-zero average correlation with the source at $x = 0$. Shaded regions in the bottom panel again depict the $1\sigma$ and $2\sigma$ uncertainties of the HMC baseline results.}
    \label{fig:fermion-correlators}
\end{figure*}

First, we confirm that each of the observables described above are measured to be consistent across sampling schemes and with HMC baseline results. Calculations of $\langle|M|\rangle$ and $\langle |\bar\psi \psi| \rangle$ using each of the generated ensembles are detailed in \Cref{tab:model-metrics} and are all consistent with the results obtained through HMC. The scalar and fermionic two-point correlators produced by the four exact Monte Carlo sampling schemes models are also consistent with the HMC baseline, as shown in \Cref{fig:scalar-correlators,fig:fermion-correlators}.

For the various sampling approaches, \Cref{tab:model-metrics} compares the autocorrelations of the magnetization and chiral condensate, as well as the Markov chain acceptance rates. Given a chain of $N$ measurements for some real-valued observable $X$, its autocorrelation function is defined as
\begin{equation}
    \Gamma_X(\tau) = \frac{1}{N-\tau} \sum_{i=1}^{N-\tau} X_i X_{i+\tau} - \left\langle X \right\rangle^2 \ ,
\end{equation}
where $\tau$ denotes the number of Markov chain steps separating the pair of measurements considered. The integrated autocorrelation time for observable $X$ is given as
\begin{equation}
    \tau^\mathrm{int}_X = \frac{1}{2} + \lim_{\tau^\mathrm{max} \rightarrow \infty} \sum_{\tau=1}^{\tau^\mathrm{max}} \frac{\Gamma_X(\tau)}{\Gamma_X(0)}\ .
\end{equation}
The sum can be truncated at a sufficiently large $\tau^\mathrm{max}$ due to the exponential suppression of $\Gamma_X(\tau)$; $1 \ll \tau^\mathrm{max} \ll N$ should be satisfied to ensure that the values of $\Gamma_X(\tau)$ are reliable. For the $\tau^\mathrm{int}$ values reported in this work, we use the Madras-Sokal windowing procedure~\cite{Madras1988} to choose a suitable $\tau^\mathrm{max}$ by identifying the earliest point where $c \tau^\textrm{int} \leq \tau^\textrm{max}$, with $c=10$. The integrated autocorrelation times $\tau^\textrm{int}_M$ and $\tau^\textrm{int}_{\bar\psi \psi}$ are given in \Cref{tab:model-metrics} together with the mean acceptance rates for all four sampling procedures.

To understand the relative performance of the four sampling approaches detailed in \Cref{tab:model-metrics}, we note that the dependence of the acceptance rate and autocorrelations on model quality is quite distinct in several of these approaches. For one, the $\phi$-marginal sampler involves an exact determinant measurement in the sampling step used for the numerical study above, which is not expected to scale efficiently. If replaced with the pseudo-marginal estimator discussed in \Cref{sec:phi-marginal-sampling}, the variance of the noisy estimates of each determinant would degrade the statistical performance achieved by even an optimally trained model and improved estimators, in particular when encountering large condition numbers. This is an obstacle to working with light fermion masses or field configuration geometries with many lattice sites (e.g.~near the thermodynamic limit) independent of the challenge of training accurate model approximations to the target distribution. Thus, the relatively higher acceptance rates and lower autocorrelation times achieved by the $\phi$-marginal sampler constructed from CPFs in this study must be contrasted against the potentially difficult scaling challenges or requirements for more precise stochastic estimators in this approach. By comparison, there is no non-trivial upper bound on the acceptance rate of the other three sampling approaches, and they will achieve an acceptance rate of 100\% when perfect model distributions are constructed.

Among these three approaches, the Gibbs sampler must also be further contrasted against the autoregressive and fully joint samplers. In particular, the remaining conditional structure of updates to $\phi$ and $\varphi$ in the Gibbs sampler results in autocorrelations even if the acceptance rate is 100\%. The magnitude of these residual autocorrelations may be small, but nevertheless puts a bound on the performance that is theoretically achievable by a Gibbs sampler, even in the asymptotic limit of perfect models of the involved distributions. Thus only joint models (either autoregressive or fully joint) can completely eliminate autocorrelations in the ideal limit of perfect models. In practice, however, the distinctions between joint models and Gibbs sampling may be minor. For example, the results presented in \Cref{tab:model-metrics} demonstrate that at the similar acceptance rates of roughly 40\%--50\% for the Gibbs and autoregressive samplers, the integrated autocorrelation times for the magnetization and condensate are similar, despite the additional autocorrelations introduced by the particular conditional structure of the Gibbs sampling scheme. The fully joint sampler shows a lower acceptance rate and greater autocorrelations, indicating that the differences are largely based on the model approximation qualities.

The particular flow-based models implemented to approximate the various distributions used for the four sampling approaches also have distinct scaling prospects. It has been found in previous work~\cite{Boyda:2020hsi} that flows based on coupling layers using convolutional networks may be easily transferred between different lattice volumes and thereby trained efficiently. This generalizability applies to the affine coupling layer implementations used for the Gibbs, autoregressive, and fully joint samplers described in this work. The CPF implementation for $\phi$-marginal sampling is also based on convolutional networks for the construction of the convex potentials, thus enabling efficient measurements of these potentials at all lattice volumes. However, in this case computing the Jacobian of the transformation to calculate $q(\phi)$ is potentially expensive, because it requires the evaluation of the Hessian of each $u_i$. Stochastic estimation of these Hessian factors may introduce additional noise in exact sampling schemes based on these particular flow architectures.

These results numerically demonstrate the effectiveness of our proposed flow models and sampling schemes. The observed performance differences cannot immediately be attributed to inherent advantages of the chosen building blocks, but may also depend strongly on the model implementation details and theory-specific characteristics. The situation may also be quite different for larger volumes and dimensions as well as other types of fields and interactions, and disentangling the effects of implementation details from asymptotic scaling properties will be the subject of future research. Furthermore, there is a large space of possible combinations of the building blocks introduced here that could be explored in future work to determine models that may have more efficient training, sampling, and scaling prospects. While an exhaustive search over this space is beyond the scope of this exploratory work, the present results serve as a guide for the design of custom flows for lattice simulations with dynamical fermions in other applications.

\section{Applicability to update-based approaches}
\label{sec:update}

\looseness=1
While the sampling schemes presented in this paper are based on the independent proposal of new field configurations (except for the Gibbs sampler; see \Cref{sec:results} for further discussion), the flow-based models defined here may also be used in methods that instead propose configuration updates, rather than independent samples. A simple way to achieve this with our architectures would be to formulate stochastic processes in the flow prior that guarantee asymptotic exactness under the target distribution, such as partial heatbath resampling, HMC, or Langevin-type algorithms, rather than independently drawing a completely new prior sample in every update step; such partial updates have previously been studied in the context of other generative models~\cite{Li2018,ZhangEWang2018Monge,Hu:2019nea} as well as trivializing map approaches~\cite{Luscher:2009eq,Engel:2011re}. In contrast to these update-based methods, direct sampling approaches have the advantage that autocorrelations in the flow-based Markov chain are in principle eliminated for an ideal model. Imperfect models, however, can still result in residual correlations caused by rejections in the Metropolis step. Whether these residual correlations from an imperfect model can outweigh the autocorrelations in corresponding update-based methods is an open question, the answer to which may also depend strongly on the model details and the specific problem under consideration.

\looseness=1
Apart from devising modified sampling schemes for the types of flows presented in this work, one may also consider defining flows that directly transform configurations in order to produce proposals for Markov chain updates. Related work on learning improved HMC-like updates includes A-NICE-MC~\cite{song2018anicemc}, its recent application to the lattice simulation of scalar $\phi^4$-theory~\cite{medvidovic2021generative}, L2HMC~\cite{levy2018generalizing}, and DLHMC~\cite{Foreman:2021ixr}, which was demonstrated to successfully mitigate topological freezing in the context of $U(1)$ lattice gauge theory in two dimensions. These approaches require the implementation of flows suitable for transforming the primary fields and conjugate momenta conditioned on each other. The flows over pseudofermion variables developed in this work can therefore be used to extend such methods to the setting of lattice field theory involving dynamical fermion fields. These insights may also inform the design of novel building blocks for the self-learning Monte Carlo method (SLMC) mentioned in the introduction, which was recently applied to non-Abelian gauge theory with dynamical fermions~\cite{Nagai:2020jar}.

\vfill\null

\section{Summary and Outlook}
\label{sec:summary}

\looseness=1
In this paper, we introduce four approaches to applying flow-based sampling to fermionic lattice field theories based on different decompositions of the joint action over bosonic and pseudofermionic fields. All approaches satisfy detailed balance and thus provide asymptotically exact sampling schemes. We further introduce several techniques to model the distributions required in these sampling approaches, including the construction of flow-based models satisfying the more complex translational symmetry group arising from the pseudofermion boundary conditions (which must be antiperiodic in time). All four sampling methods are demonstrated to successfully produce asymptotically exact samplers in a proof-of-principle application to a two-dimensional Yukawa theory.

\looseness=1
The flow-based model architectures presented here represent a selection from a large class of possible ways to model the distributions required for the four sampling methods detailed in this work. The observed relative performance of the methods in the application to the Yukawa theory provides a starting point for understanding the distinctions between these sampling approaches and architectures, but should not be considered a definitive indicator of the performance of these approaches or architectures in the context of other theories or at larger scales. Applying these methods to QCD will require future work to understand the scaling of the method with lattice volume, spacetime dimension, and with the involvement of gauge fields.

\looseness=1
Importantly, investigating the continuum limit of flow-based samplers is relevant to determine their potential to mitigate critical slowing down. This question arises with or without fermions, and empirical studies are required to understand the scaling of these methods both for fermionic and purely bosonic theories. Nonetheless, the ``building blocks'' of flows suitable for fields including pseudofermions, and the sampling strategies outlined in this work, provide the basis for developing efficient flow-based samplers for fermionic theories. 

Continued work into improved stochastic approximations of determinants~\cite{Avron:2011,Finkenrath:2012az,han2015largescale,deForcrand:2018orx,sohldickstein2020equalities,meyer2020hutch,passenheim2021variational} complements the flow-based sampling approach presented here. Such developments may be combined with our proposed flow-based sampling framework. For example, most flow-based samplers constructed in this work were designed to target the action after the application of even-odd preconditioning, a standard improvement technique for the fermionic determinant. Other preconditioners for the action may also be applied to increase the performance of flow-based samplers for these theories. Furthermore, improvements to unbiased stochastic estimates of fermionic determinants may increase the efficiency of the $\phi$-marginal sampler at scale.

\looseness=1
In summary, this work sets the stage for flow-based sampling of lattice field theories with fermions and paves the way towards an application of generative neural samplers to lattice QCD with dynamical quarks and similar problems in condensed matter theory. The next steps in this endeavor are the transfer of insights gained from the Yukawa model studied in this work to interactions between fermions and gauge fields, and the study of the scalability of the method. If they can be achieved, high quality flow-based models at the scale of state-of-the-art calculations may be able to circumvent the limitations of traditional sampling algorithms, thereby expanding the frontiers of lattice QCD and other lattice field theories.

\begin{acknowledgements}
We thank F.~Attanasio, M.~Bauer, S.~Bl\"ucher, C.~Corvalán, W.~Detmold, J.M.~Pawlowski, I.-O.~Stamatescu, N.~Strodthoff and F.P.G.~Ziegler for helpful discussions. GK, DB, DCH, and PES are supported in part by the U.S.\ Department of Energy, Office of Science, Office of Nuclear Physics, under grant Contract Number DE-SC0011090. PES is additionally supported by the National Science Foundation under EAGER grant 2035015, by the U.S.\ DOE Early Career Award DE-SC0021006, by a NEC research award, and by the Carl G and Shirley Sontheimer Research Fund. KC is supported by the National Science Foundation under the awards ACI-1450310, OAC1836650, and OAC-1841471 and by the Moore-Sloan data science environment at NYU. MSA thanks the Flatiron Institute and is supported by the Carl Feinberg Fellowship in Theoretical Physics and the James Arthur Fellowship. This work is supported by the Deutsche Forschungsgemeinschaft (DFG, German Research Foundation) under Germany's Excellence Strategy EXC 2181/1 - 390900948 (the Heidelberg STRUCTURES Excellence Cluster), the Collaborative Research Centre SFB 1225 (ISOQUANT), and the U.S.\ National Science Foundation under Cooperative Agreement PHY-2019786 (The NSF AI Institute for Artificial Intelligence and Fundamental Interactions, http://iaifi.org/). This work is associated with an ALCF Aurora Early Science Program project, and used resources of the Argonne Leadership Computing Facility, which is a DOE Office of Science User Facility supported under Contract DEAC02-06CH11357. The authors acknowledge the MIT SuperCloud and Lincoln Laboratory Supercomputing Center~\cite{reuther2018interactive} for providing HPC resources that have contributed to the research results reported within this paper.
\end{acknowledgements}

\vfill\null

\newpage
\appendix

\section{Model and training details}
\label{app:details}

\begin{table*}
    \centering
    \begin{ruledtabular}
    \begin{tabular}{m{1.3in}>{\centering\arraybackslash}p{1.0in}cccc}
      \textbf{Model parameters}   & $\phi$-Marginal  & Gibbs &   Autoregressive  & Fully Joint  \\
    \midrule
        flow layers & 3 & 16 & 12 & 12 \\ \vspace{1ex}
        convs. per layer & 4 & 3 & 10 & 6  \\ \vspace{1ex}
        number conv channels & 16 & 32  & 64  & 64 \\ \vspace{1ex}
        $\sigma_\zeta$ & $0.34$ & 0.1 & 0.34 & 0.34 \\ \vspace{1ex}
        $\sigma_\chi$  & - & 0.1 & 0.15 &  0.15 \\
        kernel size & 3 & 3 & 3 & 3 \\
        \begin{tabular}{@{}l@{}}activations \\ (inner / final) \end{tabular} & SoftPlus/- & LeakyReLU/Tanh & LeakyReLU/- & LeakyReLU/Tanh \\
    \midrule 
    \textbf{Training parameters} & & & & \\
    \midrule
        gradient steps & 500k & 30k & 200k & 50k \\ \vspace{1ex}
        batch size & 3072 & 2000 & 3072 & 3072 \\ \vspace{1ex}
        learning rate schedule 
            & \begin{tabular}{@{}c@{}}$10^{-3}$, $10^{-4}$ \\ after 80k \end{tabular} 
            & \begin{tabular}{@{}c@{}}$10^{-3}$, $10^{-5}$ \\ after 20k \end{tabular}
            & \begin{tabular}{@{}c@{}}$10^{-4}$, $2 \times 10^{-5}$ \\ after 60k, $10^{-5}$ \\ after 120k \end{tabular}
            & \begin{tabular}{@{}c@{}}$3 \times 10^{-4}$, $6 \times 10^{-5}$ \\ after 30k \end{tabular} \\
        \begin{tabular}{@{}l@{}}gradient clipping \\ (value / norm) \end{tabular} & 10/32 & -/- & 10/32 & 10/1000
    \end{tabular}
    \end{ruledtabular}
    \caption{Model and training hyperparameters for all architectures discussed in \Cref{sec:model-archs}. Further details and references are provided in \Cref{app:details}, in particular the specifics of the linear operators for the joint autoregressive model that are not listed here.}
    \label{tab:model-details}
\end{table*}

This section lists all necessary details to reproduce the flow architectures discussed in \Cref{sec:model-archs} for the application to the two-dimensional Yukawa theory studied in this work. In particular, all relevant model and training hyperparameters are given in \Cref{tab:model-details}. Additional peculiarities of the linear operator and fully joint model implementations not listed in the table are discussed below. Furthermore, we provide references for the elementary machine learning components used in the implementation and optimization. 

All models are trained using the well-known Adam optimizer \cite{kingma2017adam} with default settings. In some cases, clipping of the gradient value and norm was employed to stabilize training \cite{zhang2020gradient}. The deep neural networks providing the context functions and convex potentials in the flow architectures are implemented exclusively in terms of convolutional networks with several hidden layers and channels \cite{lecun-99}. The non-linear activation functions employed in these networks are the rectified linear unit (ReLU) \cite{agarap2019deep}, in particular the LeakyReLU variant \cite{xu2015empirical}, as well as the SoftPlus function in the case of the CPF layers, as detailed in \Cref{sec:phi-marginal-modeling}. In some cases, an additional Tanh activation is applied to the output of each network, as specified in \Cref{tab:model-details}.

For all experiments in this work using the CPF architecture, we set $w_1 = 5 \times 10^{-3}$ and $w_2 = 1$ at initialization (cf.\ \Cref{eq:cpf-conv}) and all convolutional layers use a stride of 1.

As detailed in \Cref{sec:yukawa-theory}, even-odd preconditioning of the Dirac matrix is not applied for the autoregressive architecture with linear operators, as we observe that the model gives a better approximation to the non-preconditioned action with standard lexicographic ordering. The space of possible model adjustments to make the even-odd decomposition compatible with linear operators is large, and modifications could be explored to improve the results. The conditional density $q(\varphi|\phi)$ is implemented using the composition of $128$ equivariant linear operators $\{ \mathcal{W}_k \}_{k=1}^{128}$. The $128$ linear operators are jointly defined by the stacking of a single squeezing layer breaking invariance under odd translations as explained in \Cref{app:convnet-fields}, followed by a convolutional network with periodic boundary conditions. This network features $10$ hidden layers with $64$ channels each and uses intermediate LeakyReLU activations. In total, the network has $256$ output channels, with each pair of output channels providing the values of $a$ and $b$ in the definition of one of the $128$ linear operators. The $a$ output is additionally transformed using a normalized SoftPlus function. We also find it useful to add an $L^2$ regularization loss for both outputs, with a weight of $10^{-5}$.

For the fully joint model, active components of the scalar field are transformed using its frozen components as well as the pseudofermion field. The updated scalar field together with the frozen pseudofermion components are then used to update the active pseudofermion sites.

\section{Even-odd preconditioning}
\label{app:even-odd}

Ordering lattice sites into even and odd subsets allows writing the Dirac matrix $D$ as a $2 \times 2$ block matrix of the form
\begin{equation}
\begin{aligned}
  D = \begin{pmatrix}
    m_f + g\phi_o & D_{oe} \\
    D_{eo} & m_f + g\phi_e
    \end{pmatrix}
  \equiv \begin{pmatrix}
    \mathcal{A} & \mathcal{B} \\
    \mathcal{C} & \mathcal{D}
    \end{pmatrix}\ ,
\end{aligned}
\end{equation}
where we denote the blocks as $\mathcal{A},\mathcal{B},\mathcal{C},\mathcal{D}$ for simplicity. The constant blocks $\mathcal{B} = D_{oe}$ and $\mathcal{C} = D_{eo}$ couple odd to even sites and vice versa, and $\phi_o$ and $\phi_e$ indicate the components of $\phi$ respectively associated with odd and even sites of the lattice. This form allows a more efficient stochastic approximation of the determinant by decomposing it into the determinant of either diagonal block and the associated Schur complement as
\begin{equation}\label{eq:Deo-1}
\begin{aligned}
  \det D &= \det(\mathcal{A})\, \det(\mathcal{D} - \mathcal{C} \mathcal{A}^{-1} \mathcal{B}) \\
         &= \det(\mathcal{A} \mathcal{C}^{-1} \mathcal{D} - \mathcal{B})\, \det(\mathcal{C})\ ,
\end{aligned}
\end{equation}
or, equivalently,
\begin{equation}\label{eq:Deo-2}
\begin{aligned}
  \det D &= \det(\mathcal{D})\, \det(\mathcal{A} - \mathcal{B} \mathcal{D}^{-1} \mathcal{C}) \\
         &= \det(\mathcal{D} \mathcal{B}^{-1} \mathcal{A} - \mathcal{C})\, \det(\mathcal{B})\ .
\end{aligned}
\end{equation}
Rewriting from the first to the second form in \Cref{eq:Deo-1,eq:Deo-2} ensures that the resulting expression does not involve terms that mix $\mathcal{A}$ and $\mathcal{D}$ with their respective inverses. This may lead to numerical instabilities if $m_f = 0$, which can result in ill-conditioned $\mathcal{A}$ and $\mathcal{D}$. Since $\mathcal{B}$ and $\mathcal{C}$ are constant, the terms $\det(\mathcal{B}),\det(\mathcal{C})$ drop out of the path integral and thus do not affect Metropolis-Hastings acceptance rates or gradients for optimization of flow-based models. Hence, they can be ignored for the purpose of training and sampling flow models.\footnote{If one is interested in overall estimates of $\log{Z}$, the constant contributions from these terms must then be included.}

The reduced $\nicefrac{V}{2} \times \nicefrac{V}{2}$ form of the Dirac operator makes determinant estimation significantly cheaper while keeping the additional computational overhead minimal. Half of the pseudofermion degrees of freedom completely decouple from the scalar field and can be discarded. The reduced operator partially retains the original periodic and antiperiodic boundary conditions when applied to the even or odd sub-lattices, respectively, reducing to a translation symmetry for even shifts. When utilizing affine coupling layers with a checkerboard mask, it is exactly this subset of the translational symmetry group that is preserved, and the translationally equivariant architecture is directly applicable to learning a distribution over the reduced subset of pseudofermions.

The improvement can be pushed to higher order by noting that the diagonal matrix elements of the preconditioned Dirac operator are close to unity, which makes it possible to employ an ILU preconditioning scheme~\cite{deForcrand:1996ck,Peardon:2000si}. It relies on the fact that the preconditioning matrices for the even-odd ordering step described above can be computed explicitly, which is not generally true for other ordering schemes. Though originally designed for the Wilson Dirac operator, the same procedure can be applied to the staggered fermions studied in this work. Since ILU preconditioning breaks the translation symmetry of the pseudofermion action completely, it is not directly compatible with any of our equivariant flow constructions that target the distribution of $\varphi$. However, in an experiment modeling the even-odd preconditioned $\phi$-marginal distribution using an affine coupling layer model, additional ILU preconditioning for the same architecture led to a moderately improved acceptance rate.

\section{Stochastic estimator for gradients of \texorpdfstring{$\log\det{\mathcal{M}}$}{log det M}}
\label{app:varphi-estimated-grad}

The calculation of loss gradients required for the optimization of some of the models introduced in this work requires the evaluation of gradients
\begin{equation}\label{eq:grad-logdet-M}
    \nabla_{\phi} \log\det{\mathcal{M}(\phi)}
\end{equation}
taken with respect to the field $\phi$. In general, $\mathcal{M}(\phi)$ is a positive definite matrix either arising from Dirac matrices of a pair of mass degenerate fermions as $\mathcal{M} = DD^\dagger$, or from one-flavor methods (see \Cref{sec:pseudofermions}). The calculation of the exact determinant $\det \mathcal{M}(\phi)$ may be intractable because of the scaling with the number of lattice degrees of freedom, thus a stochastic estimator is instead defined in this section to tractably evaluate \Cref{eq:grad-logdet-M}.

By assumption, $\mathcal{M}$ is a positive-definite matrix and thus the following stochastic trace estimator is applicable:
\begin{equation}\label{eq:grad-logdet-M-est}
\begin{aligned}
    \nabla \log \det \mathcal{M}(\phi) &= \nabla \Tr \log \mathcal{M}(\phi) \\
    &= \Tr\left[ \mathcal{M}(\phi)^{-1} \nabla \mathcal{M}(\phi)\right] \\
    &=\mathbb{E}_{\chi \sim e^{-\chi^\dag \chi}}[(\mathcal{M}^{-1}(\phi) \chi)^\dagger \nabla \mathcal{M}(\phi) \chi]\ .
\end{aligned}
\end{equation}
Here, the noise vector $\chi$ is assumed to be drawn from the unit-variance isotropic distribution with an appropriate number of degrees of freedom to match the dimensions of $\mathcal{M}$.

In the case of two degenerate fermionic flavors, an interesting connection can also be made to the gradient of the negative pseudofermion action (where the sign is chosen to match the positive sign of $\log \det \mathcal{M}(\phi)$). This gradient can be evaluated to be
\begin{equation}
\begin{aligned}
    \nabla (-\varphi^\dag (D D^\dag)^{-1} \varphi) &=
    \varphi^\dag (D D^\dag)^{-1} (\nabla D D^\dag) (D D^\dag)^{-1} \varphi \\
    &= \eta^{\dagger} (\nabla D D^\dag) \eta\ ,
\end{aligned}
\end{equation}
where $\eta \equiv (D D^\dag)^{-1} \varphi = (D^\dag)^{-1} \chi$, in terms of the noise vector $\chi \sim e^{-\chi^\dag \chi}$ used to generate the pseudofermion field. A short derivation shows that this is equivalent to the stochastic estimator of the two-flavor determinant,
\begin{equation}
\begin{aligned}
    &\Tr \left[ (D D^\dag)^{-1} (\nabla D D^\dag) \right] \\
    &\hspace{20pt}=
    \Tr \left[ D^{-1} (\nabla D D^\dag) (D^\dag)^{-1} \right] \\
    &\hspace{20pt}= \mathbb{E}_{\chi \sim e^{-\chi^\dag \chi}} [ ((D^\dag)^{-1} \chi)^\dag (\nabla D D^\dag) (D^\dag)^{-1} \chi] \\
    &\hspace{20pt}= \mathbb{E}_{\chi \sim e^{-\chi^\dag \chi}} [ \eta^\dag (\nabla D D^\dag) \eta]\ .
\end{aligned}
\end{equation}
This relation allows the gradient estimator to be computed using the same tools utilized for the evaluation of HMC forces with respect to the pseudofermion action.

\section{Translation-equivariant networks}

\subsection{Convolutional networks with P-fields and AP-fields}
\label{app:convnet-fields}

In \Cref{sec:translational-sym}, we introduced P-fields and AP-fields; that they form an algebra can be seen as follows. The set of P-fields is stable under linear combinations and pointwise multiplications. On the other hand, the set of AP-fields is only stable under linear combinations as the product of two AP-fields is a P-field, while the product of a P-field with an AP-field is an AP-field. In other words, the set of P-fields and AP-fields forms a superalgebra~\cite{Deligne:1999qp} under pointwise addition and multiplication. Pointwise application of a function to a P-field results in a new P-field. For AP-fields, more care is required as not all functions can be applied pointwise. Application of an odd function to an AP-field results in another AP-field, while pointwise application of an even function results in a P-field. In this appendix, we explain in more detail how properties of those fields allow us to build expressive neural networks which are equivariant under translations. This means that if $\mathcal{T} \in \mathbb{Z}^d$ is an arbitrary space-time translation and $f(\phi, \varphi) = \phi', \varphi'$ is one of our neural networks, then $f(\mathcal{T}\cdot\phi, \mathcal{T}\cdot\varphi) = \mathcal{T}\cdot\phi', \mathcal{T}\cdot\varphi'$. This discussion is not specific to $2$-dimensional spacetime, and applies for any dimension $d$.

First, convolutions can be built for both types of fields. For P-fields, this is achieved by first padding the field using periodic padding, then applying a normal convolution. For a convolution with kernel shape $2k+1$, all fields must be padded by $k$ sites in each direction. As a concrete example, assume a 1-dimensional lattice of size $5$, a P-field with values $[1, 2, 3, 4, 5]$ and a convolution kernel $[1, 1, 1]$. The padded P-field would be $[5, 1, 2, 3, 4, 5, 1]$. Applying the convolution would result in a new P-field with values $[8, 6, 9, 12, 10]$. For AP-fields, the only necessary change is to use antiperiodic padding along the time dimension, and periodic along the space dimension. Consider the 2-dimensional example
\begin{equation}
\begin{aligned}
    \begin{pmatrix}
    1 & 2 & 3 \\
    4 & 5 & 6 \\
    7 & 8 & 9 \\
    \end{pmatrix}
\end{aligned}
\end{equation}
and a $3\times 3$ kernel with all weights equal to $1$. The AP-field should be padded by $1$ site in each direction with signs applied to the temporal padding, giving
\begin{equation}
\begin{aligned}
    \begin{pmatrix}
    -9 & 7 & 8 & 9 & -7 \\
    -3 & 1 & 2 & 3 & -1 \\
    -6 & 4 & 5 & 6 & -4 \\
    -9 & 7 & 8 & 9 & -7 \\
    -3 & 1 & 2 & 3 & -1 \\
    \end{pmatrix} \ .
\end{aligned}
\end{equation}
Applying the convolution gives the transformed AP-field
\begin{equation}
\begin{aligned}
    \begin{pmatrix}
    9 & 45 & 21 \\
    9 & 45 & 21 \\
    9 & 45 & 21 \\
    \end{pmatrix} \ .
\end{aligned}
\end{equation}
The above construction of P and AP-convolutions was illustrated with only one channel, but the extension to multiple channels is straightforward.

Any non-linearity can be applied between convolutions for a P-field without spoiling translational equivariance. As mentioned above, the LeakyReLU activation function is used throughout this work. For convolutions applied to an AP-field, non-linearities used as activation functions must be restricted to odd functions, for which we choose
\begin{equation}
    \sign(\varphi)\log(1 + |\varphi|) \ .
\end{equation}

With a P-convolution, a bias can be applied along with a convolution at each step, since a bias is constant across all sites and thus transforms as a P-field. However, a traditional bias cannot be applied to the convolution of an AP-field without spoiling translational equivariance. For an AP-field $\varphi$, a bias-like operation $\varphi \rightarrow \varphi + b \sign(\varphi)$ in terms of a constant $b$ can be applied instead. To avoid potential issues with the non-differentiability of the sign function, we used a differentiable approximation given by applying $\varphi \rightarrow \varphi + b \tanh(\varphi/4)$.

All of the above constructions (P- and AP-convolutions, non-linearities and biases) are equivariant with respect to translations on the lattice. By stacking them, we create expressive translation-equivariant neural networks. Our neural networks can also jointly transform pairs of P and AP-fields. For example, we found the following transformation to work well:
\begin{verbatim}
  Input: P-field P and AP-field A
  P' = conv(P)
  A' = conv(A)
  P" = concatenate(P', |A'|)
  A" = concatenate(A', P'A')
  Output: P-field P" and AP-field A"
\end{verbatim}

The group of translational symmetries of a staggered action described in \Cref{sec:fermions-sym} only includes translations by even numbers of lattice sites. Implementing symmetries in a network that are not symmetries of the target function restricts the expressivity and may make it difficult or impossible to represent an effective approximation of the target function by the network. For the models targeting the staggered fermion action in the study described in the main text, we avoided encoding translational symmetry by an odd number of sites by explicitly breaking equivariance with respect to odd translations. For example, to break the symmetry by odd translations along the first dimension of a field $x$, we fold its even and odd indices along the channel dimension; this doubles its number of channels while halving the number of points along the first dimension. We then apply a convolution with stride 1 and kernel size 1, which mixes all the channels. Finally, we split the channels in two and fold them back along the first dimension to get a new field with the same lattice size as $x$. This approach mirrors the `squeezing' operation applied in Real NVP flows~\cite{dinh2017density}.

\subsection{Explicit symmetrization by group averages}
\label{app:symmetrization}

Equivariance of neural networks under discrete symmetry groups can also be achieved by explicitly averaging over the group. For discrete translations with antiperiodic boundary conditions in particular, this approach avoids the restriction to odd activation functions and vanishing biases, but requires an increase of the computational effort proportional to the timelike extension of the lattice.

Let $\mathcal{T}^a_{\vec{x},t} \in \mathbb{Z}^d$ denote a translation by $(\vec{x},t)$ where antiperiodic boundary conditions are applied in time and periodic boundary conditions are applied in space, and let $\mathcal{T}^p_{\vec{x},t}$ denote a translation by $(\vec{x},t)$ with periodic boundary conditions for all directions. The action of $\mathcal{T}^a$ and $\mathcal{T}^p$ is the same along all spatial dimensions, and we define $\mathcal{T}_{\vec{x}} = \mathcal{T}^p_{\vec{x}, 0} = \mathcal{T}^a_{\vec{x}, 0}$. For simplicity we now restrict to working with a two-dimensional $L \times L$ lattice with coordinates $(\vec{x},t)$, but the following construction immediately generalizes to higher dimensions and non-symmetric lattices. Both P-fields and AP-fields are maps from $\mathbb{Z}_{L} \times \mathbb{Z}_{L}$ to $\mathbb{R}^c$, where $c$ is a number of channels.
Under lattice translations, P-fields are acted upon by $\mathcal{T}^p$, while AP-fields are acted upon by $\mathcal{T}^a$.

Consider a function $f$ that maps the pair $(\phi, \varphi)$ of a P-field and an AP-field to another field $f(\phi, \varphi)$. Assume that the output $f(\phi, \varphi)$ transforms with periodic boundary conditions along the space dimension, that is:
\begin{equation}\label{eq:f-equivariance}
    f(\mathcal{T}_{\vec{x}} \phi, \mathcal{T}_{\vec{x}} \varphi) = \mathcal{T}_{\vec{x}} f(\phi, \varphi)\ .
\end{equation}
Using averaging, we will now construct two maps $u$ and $v$ with the transformation properties:
\begin{equation}\label{eq:s-equivariance-prop}
    u(\mathcal{T}^p_{\vec{x}, t}\phi, \mathcal{T}^a_{\vec{x}, t}\varphi) = \mathcal{T}^p_{\vec{x}, t}u(\phi, \varphi)
\end{equation}
\begin{equation}\label{eq:t-equivariance-prop}
    v(\mathcal{T}^p_{\vec{x}, t}\phi, \mathcal{T}^a_{\vec{x}, t}\varphi) = \mathcal{T}^a_{\vec{x}, t}v(\phi, \varphi) \ .
\end{equation}
We define\footnote{Note that if $f$ is odd, then $u$ below will be forced to be independent of $\varphi$. This can be avoided by either using non-odd non-linearities, or by having non-zero biases.}
\begin{equation}
    u(\phi, \varphi) = \frac{1}{2L}\sum_{n=0}^{2L-1}\mathcal{T}^p_{0,-n}f(\mathcal{T}^p_{0,n}\phi, \mathcal{T}^a_{0,n}\varphi)
\end{equation}
and
\begin{equation}
    v(\phi, \varphi) = \frac{1}{2L}\sum_{n=0}^{2L-1}\mathcal{T}^a_{0,-n}f(\mathcal{T}^p_{0,n}\phi, \mathcal{T}^a_{0,n}\varphi) \ .
\end{equation}

We now wish to prove that the transformation properties in \Cref{eq:s-equivariance-prop} and \Cref{eq:t-equivariance-prop} apply to these definitions. The proof is roughly the same for both cases,\footnote{The proof is actually a particular instance of a more general property: if $\pi_1$ and $\pi_2$ are representations of a finite group $G$ on vector spaces $V$ and $W$, and if $f:V\rightarrow W$ is any map between these spaces, then $\frac{1}{|G|}\sum_{g\in G}\pi_2(g)^{-1}f(\pi_1(g))$ is an equivariant map from $V$ to $W$.} so we will only write it for \Cref{eq:s-equivariance-prop}:
\begin{equation}
\begin{aligned}
    u(\mathcal{T}^p_{\vec{x}, t}\phi, \mathcal{T}^a_{\vec{x}, t}\varphi) &= \frac{1}{2L}\sum_{n=0}^{2L-1}\mathcal{T}^p_{0,-n}f(\mathcal{T}^p_{0,n}\mathcal{T}^p_{\vec{x}, t}\phi, \mathcal{T}^a_{0,n}\mathcal{T}^a_{\vec{x}, t}\varphi)  \\
    &= \frac{1}{2L}\sum_{n=0}^{2L-1}\mathcal{T}^p_{0,-n}f(\mathcal{T}_{\vec{x}}\mathcal{T}^p_{0,n+t}\phi, \mathcal{T}_{\vec{x}}\mathcal{T}^a_{0,n+t}\varphi)  \\
    &= \frac{1}{2L}\sum_{n=0}^{2L-1}\mathcal{T}^p_{0,-n+t}\mathcal{T}_{\vec{x}}f(\mathcal{T}^p_{0,n}\phi, \mathcal{T}^a_{0,n}\varphi)\label{eq:s-equivariance-step}\\
    &= \mathcal{T}^p_{\vec{x}, t}u(\phi, \varphi) \ .
\end{aligned}
\end{equation}
\Cref{eq:s-equivariance-step} was obtained using the change of variables $n\rightarrow n-t$ and the equivariance property given in \Cref{eq:f-equivariance}.

\looseness=1
The functions $u$ and $v$ may be used to define equivariant affine coupling layers for the construction of equivariant flows. To achieve equivariance, the underlying function $f$ needs to be evaluated $2L$ times instead of once, hence the aforementioned increase in computational cost. Since the masked affine couplings employed in this work already restrict the translational equivariance to multiples of two, one may also consistently use only every second term in the sums defining $u$ and $v$ without breaking the symmetry further, implying a factor $L$ increase of the cost instead of $2L$. Still, the additional computational requirements are significant compared to the approach detailed in \Cref{app:convnet-fields}, and for large-scale implementations one may have to partially trade equivariance against efficiency by excluding more terms from the sums.

\newpage
\bibliographystyle{utphys}
\bibliography{main}

\end{document}